\documentclass[prd,aps,preprint,tightenlines,onecolumn,superscriptaddress,nofootinbib,eqsecnum,preprintnumbers,longbibliography,12pt]{revtex4-1}
\pdfoutput=1
\usepackage{epsfig,amsfonts,amsmath,mathrsfs,latexsym,amssymb,graphicx,ifpdf,slashed,color,url,cancel,enumerate,braket,xspace,colordvi}
\usepackage[T1]{fontenc}
\usepackage{tikz}
\usepackage[compat=1.0.0]{tikz-feynman}
\usepackage[colorlinks=true,allcolors=purple]{hyperref}
\usepackage[capitalize]{cleveref}

\begin{document}
\preprint{MI-HET-868}

\title{Radiative Correction from Secret Neutrino Interactions and Implications for Neutrino-Scattering Experiments}

\author{Saeid Foroughi-Abari}
\affiliation{Department of Physics, Carleton University, Ottawa, Ontario K1S 5B6, Canada}

\author{Kevin J. Kelly}
\affiliation{Department of Physics and Astronomy, Mitchell Institute for Fundamental Physics and Astronomy,
Texas A\&M University, College Station, TX 77843, USA}

\author{Yue Zhang}
\affiliation{Department of Physics, Carleton University, Ottawa, Ontario K1S 5B6, Canada}

\date{\today}

\begin{abstract}
New, neutrinophilic mediators are one potential extension beyond the Standard Model of particle physics.
Often, studies of neutrinophilic mediator consist of searching for direct evidence of its production and/or its tree-level virtual effect for generating strong neutrino self-interaction.
In this work, we focus instead on the fact that such new mediators \textit{also} lead to deviations in neutrino-matter scattering via radiative corrections.
With a mediator mass well below the electroweak scale, these effects are potentially observable in a variety of contexts, including coherent elastic neutrino-nucleus scattering (CEvNS), neutrino deeply-inelastic scattering ($\nu$DIS), and neutrino-electron scattering (e.g., at Borexino).
Additionally, such effects lead to new contributions to the $Z$-boson decay width and to non-standard neutrino interactions relevant for long-baseline oscillation experiments.
We explore all of these scenarios in some depth, building on the rich phenomenology associated with neutrinophilic mediators.
\end{abstract}

{
\let\clearpage\relax
\maketitle
}

\begingroup
    \let\clearpage\relax
    \tableofcontents
\endgroup

\section{Introduction}
State-of-the-art terrestrial and astrophysical explorations leave ample space for self-interactions among the active neutrinos, even much more strongly than predicted by the Standard Model. 
Novel neutrino self-interaction is well motivated. 
It can account for the relic abundance of sterile neutrino dark matter produced via an oscillation mechanism in the early universe~\cite{DeGouvea:2019wpf, Kelly:2020pcy, Kelly:2020aks} and remain consistent with small scale structure constraints~\cite{An:2023mkf}. 
Analyses of the cosmic microwave background and large-scale structure also hint for the role of strong neutrino self-interactions in the early universe~\cite{Kreisch:2019yzn, Park:2019ibn, Das:2020xke, He:2023oke, Camarena:2023cku, Das:2023npl, Camarena:2024zck, Pal:2024yom, Racco:2024lbu}.
In recent years, there has been a wealth of literature investigating the fundamental physics behind neutrino self-interactions and charting the road map for upcoming searches~\cite{Bardin:1970wq, Barger:1981vd, Beacom:2002cb, Laha:2013xua, Ioka:2014kca, Ng:2014pca, Forastieri:2015paa, Picoreti:2015ika, Shoemaker:2015qul, Heurtier:2016otg, Das:2017iuj, Huang:2017egl, Berryman:2018ogk, Blum:2018ljv, Brune:2018sab, Kelly:2019wow, Forastieri:2019cuf, Funcke:2019grs, Blinov:2019gcj, deGouvea:2019qaz, Shalgar:2019rqe, EscuderoAbenza:2020cmq, Bustamante:2020mep, Grohs:2020xxd, Brdar:2020nbj, Lyu:2020lps, Deppisch:2020sqh, Reddy:2021rln, Esteban:2021tub, Picoreti:2021yct, Ge:2021lur, Chichiri:2021wvw, Kelly:2021mcd, Smirnov:2022sfo, Berryman:2022hds, deGouvea:2022cmo, Das:2022xsz, Akita:2022etk, Chang:2022aas, Chen:2022kal, Fiorillo:2022cdq, Coyle:2022bwa, Doring:2023vmk, Sandner:2023ptm, Hyde:2023eph, Li:2023puz, Fiorillo:2023ytr, Fiorillo:2023cas, Wu:2023twu, Graf:2023dzf, Telalovic:2024cot, Huang:2024tbo, Dev:2024ygx, Bai:2024kmt, Zhang:2024meg, Adhikary:2024tvl, Liu:2024ywd, Foroughi-Abari:2025upe, Wang:2025qap, Montefalcone:2025unv, Leal:2025eou, He:2025bex, Das:2025asx, Dev:2025tdv, Kling:2025zsb, Montefalcone:2025mbg, Vogel:2025gan, Montefalcone:2025ibh}.

The leading model for strong neutrino self-interactions is to introduce a light gauge-singlet scalar $\phi$ with a Yukawa coupling to a pair of neutrinos.
This is a simplified model and the Yukawa interaction can originate from a gauge-invariant, higher-dimensional operator of the form $(LH)^2\phi$~\cite{Berryman:2018ogk}.
This operator serves as the starting point of many analyses studying the phenomenological aspects of a light neutrinophilic scalar.

The necessity of invoking a renormalizable ultraviolet (UV) completion for this dimension-six operator was recently pointed out in~\cite{Zhang:2024meg}. 
Any heavy particle(s) integrated out leading to the $(LH)^2\phi$ operator plays a crucial role in rendering electroweak processes free from ultraviolet (UV) divergences at the one-loop level.
In this work, we present detailed calculations of the radiative corrections induced in electroweak couplings involving neutrinos, in the context of a scalar $SU(2)_L$ triplet extension of the Standard Model (detailed in~\cref{sec:model}).
This UV-completes the self-interacting neutrino model and generates nontrivial behavior in a number of commonly observed neutrino-scattering processes.

Our result of radiative corrections in~\cref{sec:radiativec} will be useful in making precise predictions for these processes in order to confront experimental findings.
We apply this to several precision Standard Model tests involving neutrino neutral-current scattering.
The impact on current and future weak-mixing-angle measurements using neutrino scattering is discussed in~\cref{sec:WMA}.
We revisit and sharpen the discussion of the invisible $Z$-boson decay and its constraint on such neutrinophilic mediator models in~\cref{sec:Z}.
In~\cref{sec:NSI}, we derive a new, leading constraing on $\nu_\tau$-philic self-interactions for mediator masses above approximately $20$~MeV and below a few GeV.
We explore the process of solar neutrinos scattering elastically off electrons and Borexino's measurement of this process in~\cref{sec:solar}.
Finally, we offer some concluding remarks and discussion in~\cref{sec:discussion}.

%%%%%%%%%%%%%%%%%%%%%%%%%%%%%%%%%%%%%%%%%%%%%%%%%%%%%%%%%%%%%%%%%%%%%%%%%%%
\section{Neutrinophilic scalar model and UV completion}\label{sec:model}
%%%%%%%%%%%%%%%%%%%%%%%%%%%%%%%%%%%%%%%%%%%%%%%%%%%%%%%%%%%%%%%%%%%%%%%%%%%
The low-energy simplified model for neutrino self-interactions involving a neutrinophilic scalar mediator $\phi$ has the following interaction Lagrangian,
\begin{equation}\label{eq:Lsim}
\mathcal{L} = \sum_{\alpha,\beta=e,\mu,\tau}\frac{\lambda_{\alpha \beta}}{2} \, \bar\nu_\alpha \nu_\beta^c \, \phi + \text{h.c.} \ ,
\end{equation}
where $\nu_\alpha$ represents the active neutrinos that belong to the lepton doublets under $SU(2)_L$, $\nu^c = i \gamma^2\nu^*$ denotes the right-handed antineutrino field, and $\phi$ is a Standard-Model (SM) gauge-singlet with no direct coupling to the other particles.
For concreteness, we assume $\phi$ to be a complex scalar. This assumption is not necessary; the same result holds for the case where $\phi$ is real.
In this work, we explore the mass range of $\phi$ below the electroweak (EW) scale.
For the flavor-dependent coupling parameters $\lambda_{\alpha \beta}$, popular literature choices include the flavor-universal/diagonal case ($\lambda_{\alpha\beta} = \lambda \delta_{\alpha\beta}$), and the $\nu_\tau$-specific case ($\lambda_{\alpha\beta} = \lambda \delta_{\alpha\tau} \delta_{\beta\tau}$), where $\delta$ is the Kronecker delta function.

The interaction Lagrangian in~\cref{eq:Lsim} can be obtained from gauge-invariant operators
\begin{equation}\label{eq:Oeff}
(\bar L_\alpha i \sigma_2 H^*) (H^\dagger i \sigma_2 L_\beta^c) \, \phi + \text{h.c.} \ ,
\end{equation}
where $L_\alpha = (\nu_\alpha, \ell_\alpha)^T$ is a SM lepton doublet, and $L^c_\beta = (\nu_\beta^c, \ell_\beta^c)^T$.
Inside each bracket, the antisymmetric tensor $i \sigma_2$ operates in the $SU(2)_L$ space.

To facilitate the discussion of radiative corrections in the presence of a light $\phi$, we present a UV-complete model for the effective operator in~\cref{eq:Oeff} which extends the SM with $\phi$ and an $SU(2)_L$-triplet scalar boson $T$ with hypercharge $-2$,
\begin{equation}
T = \begin{pmatrix}
    T^-/\sqrt{2} & T^0 \\
    T^{--} & - T^-/\sqrt{2}
\end{pmatrix} \ ,
\end{equation} 
whose two indices transform as $2\otimes \bar 2$ under $SU(2)_L$.
Its kinetic term takes the form ${\rm Tr}[(D^\mu T)^\dagger (D_\mu T)]$ and the covariant derivative is
\begin{equation}
D_\mu T = \partial_\mu T + \frac{g}{2} \left[ W_\mu^a \sigma^a , T\right] - g' B_\mu T \ ,
\end{equation} 
where $W^a_\mu$ and $B_\mu$ are the gauge bosons of $SU(2)_L$ and $U(1)_Y$, respectively, and $\sigma^a$ are Pauli matrices.
The gauge-invariant, renormalizable interaction Lagrangian for $T$ reads
\begin{equation}\label{eq:UV}
\mathcal{L} = - \sum_{\alpha,\beta=e,\mu,\tau} \frac{y_{\alpha \beta}}{2} \,\bar L_\alpha T i\sigma_2 L_\beta^c  + y' H^T i\sigma_2 T H \phi^* + \text{h.c.} \ .
\end{equation}
The neutral component of the triplet, $T^0$, only couples to neutrinos.

After EW symmetry breaking, $\langle H\rangle = (0, v/\sqrt{2})^T$, a mass-mixing term is generated between $T^0$ and $\phi$ (both are complex scalars),
\begin{equation}
-\begin{pmatrix}
    \phi^* & T^{0*}
\end{pmatrix}
\begin{pmatrix}
    \mu_\phi^2 & y' v^2/2 \\
    y'^* v^2/2 & M^2
\end{pmatrix}
\begin{pmatrix}
    \phi \\ T^0
\end{pmatrix} \ ,
\end{equation}
where $\mu_\phi^2$ and $M^2$ are the mass-squared parameters for $\phi$ and $T$ from their potentials.
In the large-$M$ limit, the mixing angle to diagonalize this matrix is approximately
\begin{equation}\label{eq:theta}
\sin\theta \simeq \frac{|y'| v^2}{2M^2} \ .
\end{equation}
In the mass basis, the physical states $\hat\phi$ and $\hat T^0$ are linear combination of the original $\phi$ and $T^0$.
Their masses are
\begin{equation}
m_{\hat \phi}^2 \simeq \mu_\phi^2 - \frac{|y'|^2 v^4}{4M^2} \ , \quad  \quad  
m_{\hat T^0} \simeq M \ .
\end{equation}
Through the mixing, $\hat\phi$ obtains a Yukawa coupling with the neutrino bilinear operator, of the form~\cref{eq:Lsim}, with the relation
\begin{equation}
\lambda_{\alpha\beta} = y_{\alpha\beta} \sin\theta \ .
\end{equation}
In this concrete UV completion, the coupling $\lambda_{\alpha\beta}$ is proportional to the $\phi$-$T^0$ mixing $\sin\theta$ which can serve as a small expansion parameter in the large-$M$ limit.
For small $\theta$, $\hat\phi$ primarily consists of the original gauge-singlet $\phi$ with a small mixture of $T^0$.

In this model, we assume that neither $T^0$ nor $\phi$ has a vacuum expectation value, thus the model considered here plays the sole role of generating novel neutrino self-interaction. Connections to the origin of neutrino masses would require further model building and receive additional theoretical constraints that are beyond the scope of this work.

We present the relevant gauge and Yukawa Feynman rules for $\hat\phi$ and $\hat T^0, T^-, T^{--}$ from the triplet in~\cref{appA}, which are useful for the radiative correction calculation explored in the next section.
For simplicity, we will suppress the hat over the physical fields for the remainder of this work.

\section{Radiative correction to Neutrino Electroweak Couplings}
\label{sec:radiativec}

In this section, we derive radiative corrections to the neutrino EW couplings assuming that $\phi$ is the only new particle below the electroweak scale.
The triplet mass scale $M$ lies above the EW scale, allowing us to integrate out $T$.
The discussions here provide detailed calculations behind the arguments made in Ref.~\cite{Zhang:2024meg} and generalize the results therein.

The EW Lagrangian of interest here is~\cite{ParticleDataGroup:2024cfk}
\begin{equation}
    \mathcal{L} = - (g_Z)_{\alpha\beta} \bar \nu_\alpha \gamma^\mu \mathbb{P}_L \nu_\beta - (g_W)_{\alpha\beta} \bar \ell_\alpha \gamma^\mu \mathbb{P}_L \nu_\beta W^-_\mu + {\rm h.c.} \ ,
\end{equation}
where $\mathbb{P}_L = (1-\gamma_5)/2$, and $\alpha, \beta = e, \mu, \tau$ are flavor indices.
At tree level, the gauge couplings are only nonzero for $\alpha=\beta$ and 
\begin{equation}
g_Z^0 = \frac{g}{2\cos\theta_W} \ , \quad g_W^0=\frac{g}{\sqrt2} \ . 
\end{equation}
With the above definitions of $g$ and $v$, the Fermi constant is given by $G_F =\sqrt{2}g^2/(8M_W^2) = (\sqrt{2}v^2)^{-1}$.

Next, we calculate the radiative corrections to the gauge couplings $g_Z$ and $g_W$ from neutrino self-interaction. In the simplified model Eq.~\eqref{eq:Lsim}, there are two relevant one-loop diagrams
\begin{align}\label{eq:diagrams1}
\begin{tikzpicture}
\begin{feynman}
\vertex [label=above:\(Z\)] (a) at (0,0);
\vertex (b) at (0.7,0);
\vertex (c) at (1.57,0.5);
\vertex (d) at (1.57,-0.5);
\vertex [label=right:\(\nu_\alpha\)] (e) at (2.27,0.5);
\vertex [label=right:\(\nu_\beta\)] (f) at (2.27,-0.5);
\feynmandiagram [inline=(a.base)] {
(a)  -- [photon, momentum'=\(q\)] (b),
(b) -- [fermion, edge label=\(\nu^c_\gamma\)] (c),
(d) -- [charged scalar, edge label'=\(\phi\)] (c),
(d) -- [fermion, edge label=\(\nu^c_\gamma\)] (b),
(c) -- [fermion] (e),
(d) -- [anti fermion] (f),
};
\end{feynman}
\end{tikzpicture} 
\hspace{0.5cm}
\raisebox{0.8cm}{
$\equiv -i g_{Z}^0 \Gamma^{(1)}_{Z\bar\nu_\alpha\nu_\beta}(q^2, m_\phi^2)$ \ ,
} \\
\feynmandiagram [inline=(a.base), horizontal=a to d] {
  i1 [particle=\(\nu_\beta\)] -- [fermion, momentum'=\(p\)] a -- [fermion, edge label'=\(\nu^c_\gamma\)] b -- [fermion] i2 [particle=\(\nu_\alpha\)],
  a -- [charged scalar, half left, looseness=1.5, edge label=\(\phi\)] b,
}; 
\equiv -i\Sigma_{\bar\nu_\alpha \nu_\beta}(p, m_\phi^2) \ , \nonumber
\end{align} 
where $\Gamma_{Z\nu\bar\nu}(q^2, m_\phi^2)$ and $\Sigma_\nu(p, m_\phi^2)$ are the amputated vertex correction and neutrino self-energy diagrams. 
$q^\mu$ ($p^\mu$) is the momentum running through the incoming $Z$-boson (incoming/outgoing neutrino legs).
The loops are equal to
\begin{eqnarray}\label{eq:loopfunctions1}
&&\Gamma_{Z\bar\nu_\alpha\nu_\beta}^{(1)}(q^2, m_\phi^2) = -\frac{\lambda_{\alpha\gamma}\lambda^*_{\beta\gamma}}{32\pi^2} 
\left\{ \frac{2}{\varepsilon} - 1 + 2\int_0^1 dx \int_0^{1-x}dy \right. \nonumber\\
&&\hspace{2.5cm}\left.\times\left[\ln \left(\frac{\mu^2}{(1-x-y)m_\phi^2-xy q^2}\right) + \frac{x y q^2}{(1-x-y)m_\phi^2-xy q^2} \right] \right\} \ ,\nonumber\\
&&\Sigma_{\bar\nu_\alpha\nu_\beta}(p, m_\phi^2) = -\frac{\lambda_{\alpha\gamma}\lambda^*_{\beta\gamma}}{32\pi^2} \left( \frac{2}{\varepsilon} + \frac{1}{2} + \log \frac{\mu^2}{m_\phi^2}\right) \cancel{p} \ ,
\end{eqnarray}
where $2/\varepsilon = 2/(4-d)-\gamma_E + \ln(4\pi)$ is used to regularize the UV divergence in dimensional regularization, $\gamma_E=0.577$ is Euler's constant, and $\mu$ is the renormalization scale.
The sum over flavor index $\gamma$ is implicit.

Because $\phi$ is neutrinophilic by definition, there is no one-loop correction to the charged-current vertex in the simplified model.
If we take the above loop functions to compute the renormalized gauge couplings, the result will be UV divergent~\cite{Zhang:2024meg}.
It is a manifestation of the incompleteness of the simplified model which fails to respect SM gauge invariance.
To solve this problem, we must resort to a UV-complete model, {\it e.g.}, the $SU(2)_L$ triplet scalar extension presented above.

In the UV-complete model, it is the triplet $T$ that couples to the lepton doublets at the most fundamental level.
It is useful to first comment on the renormalization of gauge couplings in the $\theta=0$ limit.
The only effect of $T$ on the running of gauge couplings occurs through the gauge-boson vacuum polarization diagrams.
As a result, the corresponding $\beta$ functions do not depend on any Yukawa couplings at one-loop level.
On the other hand, the contributions from vertex corrections and neutrino self-energy sum up to be UV finite by virtue of the Ward identity.
For sufficiently heavy $T$, the finite correction is suppressed by the ratio of $q^2$ to the triplet mass-squared ($\sim q^2/M^2$) in the large-$M$ limit, as required by the decoupling theorem. 

Switching on the $\phi$-$T^0$ mixing will not affect the above UV finiteness which is about physics at much shorter distances.
More precisely, the $\phi$-neutrino coupling in the simplified model is generated via the mixing between $\phi$ and $T^0$ after EW symmetry breaking.
Both diagrams in Eq.~\eqref{eq:diagrams1} are proportional to $\sin^2\theta$ (note that $\lambda_{\alpha\gamma}\lambda^*_{\beta\gamma}=y_{\alpha\gamma}y^*_{\beta\gamma}\sin^2\theta$), where $\theta$ is the $\phi$-$T^0$ mixing parameter introduced in~\cref{eq:theta}.
There are additional Feynman diagrams involving $\phi$ and $T^0$ that contribute at the same $y^2\sin^2\theta$ order and must be taken into account.

Before proceeding, it is crucial to remark that in this work we are interested in low-energy neutrino scattering processes with $|q^2| \ll v^2$.
%The present experimental sensitivities to new physics are of percent level.
With this in mind, we further assume the hierarchy 
\begin{align}\label{eq:hierarchy}
\frac{|q^2|}{M^2} \ll \frac{v^4}{M^4} \sim \sin^2\theta \ ,
\end{align}
which implies that the triplet mass cannot be too far above the EW scale.
Under the assumption of~\cref{eq:hierarchy}, the triplet contribution in the $\theta=0$ limit is subdominant to the $\mathcal{O}(\sin^2\theta)$ results that will be derived in~\cref{eq:6,eq:7} below.

In the UV-complete model, the $\phi$-$T^0$ mixing leads to additional vertex-correction diagrams that contribute at $\sin^2\theta$ order,
\begin{align}\label{eq:diagrams2}
\begin{tikzpicture}
\begin{feynman}
\vertex [label=above:\(Z\)] (a) at (0,0);
\vertex (b) at (0.7,0);
\vertex (c) at (1.57,0.5);
\vertex (d) at (1.57,-0.5);
\vertex [label=right:\(\nu_\alpha\)] (e) at (2.27,0.5);
\vertex [label=right:\(\nu_\beta\)] (f) at (2.27,-0.5);
\feynmandiagram [inline=(a.base)] {
(a)  -- [photon, momentum'=\(q\)] (b),
(b) -- [charged scalar, edge label=\(T^0\)] (c),
(d) -- [fermion, edge label'=\(\nu^c_\gamma\)] (c),
(d) -- [charged scalar, edge label=\(\phi\)] (b),
(c) -- [fermion] (e),
(d) -- [anti fermion] (f),
};
\end{feynman} 
\end{tikzpicture}
\hspace{1cm} \raisebox{0.8cm}{+} \hspace{0.5cm}
\begin{tikzpicture}
\begin{feynman}
\vertex [label=above:\(Z\)] (a) at (0,0);
\vertex (b) at (0.7,0);
\vertex (c) at (1.57,0.5);
\vertex (d) at (1.57,-0.5);
\vertex [label=right:\(\nu_\alpha\)] (e) at (2.27,0.5);
\vertex [label=right:\(\nu_\beta\)] (f) at (2.27,-0.5);
\feynmandiagram [inline=(a.base)] {
(a)  -- [photon, momentum'=\(q\)] (b),
(b) -- [charged scalar, edge label=\(\phi\)] (c),
(d) -- [fermion, edge label'=\(\nu^c_\gamma\)] (c),
(d) -- [charged scalar, edge label=\(T^0\)] (b),
(c) -- [fermion] (e),
(d) -- [anti fermion] (f),
};
\end{feynman} 
\end{tikzpicture}
\hspace{0.5cm}
\raisebox{0.8cm}{
$\equiv -i g_{Z}^0 \Gamma^{(2)}_{Z\bar\nu_\alpha\nu_\beta}(q^2, m_\phi^2, M^2)$ \ ,
}\\
\begin{tikzpicture}
\begin{feynman}
\vertex [label=above:\(W^-\)] (a) at (0,0);
\vertex (b) at (0.7,0);
\vertex (c) at (1.57,0.5);
\vertex (d) at (1.57,-0.5);
\vertex [label=right:\(\ell_\alpha^-\)] (e) at (2.27,0.5);
\vertex [label=right:\(\nu_\beta\)] (f) at (2.27,-0.5);
\feynmandiagram [inline=(a.base)] {
(a)  -- [charged boson, momentum'=\(q\)] (b),
(b) -- [charged scalar, edge label=\(T^-\)] (c),
(d) -- [fermion, edge label'=\(\nu^c_\gamma\)] (c),
(d) -- [charged scalar, edge label=\(\phi\)] (b),
(c) -- [fermion] (e),
(d) -- [anti fermion] (f),
};
\end{feynman} 
\end{tikzpicture} 
\hspace{0.5cm}
\raisebox{0.8cm}{
$\equiv -i g_{W}^0 \Gamma_{W\bar\ell_\alpha\nu_\beta}(q^2, m_\phi^2, M^2)$ \ ,} \nonumber
\end{align}
The corresponding loop functions are
\begin{eqnarray}\label{eq:loopfunctions2}
\Gamma_{Z\bar\nu_\alpha\nu_\beta}^{(2)}(q^2, m_\phi^2, M^2) &\simeq& \frac{\lambda_{\alpha\gamma}\lambda^*_{\beta\gamma}}{8\pi^2} 
\left\{ \frac{2}{\varepsilon} + 2\int_0^1 dx \int_0^{1-x}dy \ln \left(\frac{\mu^2}{xM^2 + y m_\phi^2-xy q^2}\right) \right\}  \nonumber\\
\Gamma_{W\bar\ell_\alpha\nu_\beta}(q^2, m_\phi^2, M^2)&\simeq& \frac{1}{4}\Gamma_{Z\nu\bar\nu}^{(2)}(q^2, m_\phi^2, M^2) \ . 
\end{eqnarray}
Hereafter, we use  a universal scale $M$ to denote the scalar triplet mass.
The mass difference among its components is tightly constrained by EW precision observables~\cite{Melfo:2011nx}.
In the second line, the relative factor $1/4$ can be understood from the number of Feynman diagrams in Eq.~\eqref{eq:loopfunctions2} and the triplet gauge and Yukawa Feynman rules given in~\cref{appA}.

The above is not yet the entire story.
For the neutral-current vertex, we also need to take into account another class of contribution from the same set of the above Feynman diagrams,~\cref{eq:diagrams1,eq:diagrams2}, but with $\phi$ replaced by $T^0$ and the corresponding Yukawa coupling to neutrino is $y$ instead of $\lambda$.
The counterparts of $\Gamma^{(1)}_{Z\bar\nu_\alpha\nu_\beta}$ and $\Sigma_{\bar\nu_\alpha \nu_\beta}$ are proportional to $y_{\alpha\gamma}y^*_{\beta\gamma}\cos^2\theta$ instead of $\lambda_{\alpha\gamma}\lambda^*_{\beta\gamma}$.
The counterpart of $\Gamma^{(2)}_{Z\bar\nu_\alpha\nu_\beta}$ is proportional to $y_{\alpha\gamma}y^*_{\beta\gamma}\cos^4\theta$.
As mentioned earlier, these contributions exist in the UV-complete theory in the absence of $\phi$-$T^0$ mixing.
For $\theta=0$, the decoupling theorem applies to the sum of these diagrams along with the following ones involving $T^\pm$,
%but not $T^0$ (thus $\theta$-independent),
\begin{align}
\begin{tikzpicture}
\begin{feynman}
\vertex [label=above:\(Z\)] (a) at (0,0);
\vertex (b) at (0.7,0);
\vertex (c) at (1.57,0.5);
\vertex (d) at (1.57,-0.5);
\vertex [label=right:\(\nu_\alpha\)] (e) at (2.27,0.5);
\vertex [label=right:\(\nu_\beta\)] (f) at (2.27,-0.5);
\feynmandiagram [inline=(a.base)] {
(a)  -- [photon] (b),
(b) -- [fermion, edge label=\(\ell^+_\gamma\)] (c),
(d) -- [charged scalar, edge label'=\(T^-\)] (c),
(d) -- [fermion, edge label=\(\ell^+_\gamma\)] (b),
(c) -- [fermion] (e),
(d) -- [anti fermion] (f),
};
\end{feynman}
\end{tikzpicture} 
\hspace{2cm}
\raisebox{0.15cm}{\begin{tikzpicture}
\begin{feynman}
\vertex [label=above:\(Z\)] (a) at (0,0);
\vertex (b) at (0.7,0);
\vertex (c) at (1.57,0.5);
\vertex (d) at (1.57,-0.5);
\vertex [label=right:\(\nu_\alpha\)] (e) at (2.27,0.5);
\vertex [label=right:\(\nu_\beta\)] (f) at (2.27,-0.5);
\feynmandiagram [inline=(a.base)] {
(a)  -- [photon] (b),
(b) -- [charged scalar, edge label=\(T^-\)] (c),
(d) -- [fermion, edge label'=\(\ell^+_\gamma\)] (c),
(d) -- [charged scalar, edge label=\(T^-\)] (b),
(c) -- [fermion] (e),
(d) -- [anti fermion] (f),
};
\end{feynman} 
\end{tikzpicture}}
\hspace{2cm}
\raisebox{1cm}{
\feynmandiagram [inline=(a.base), horizontal=a to d] {
  i1 [particle=\(\nu_\beta\)] -- [fermion] a -- [fermion, edge label'=\(\ell^+_\gamma\)] b -- [fermion] i2 [particle=\(\nu_\alpha\)],
  a -- [charged scalar, half left, looseness=1.5, edge label=\(T^-\)] b,
};} 
\end{align}
The sum is finite and in the large-$M$ limit is negligible for low-$q^2$ processes under our assumption in~\cref{eq:hierarchy}. 
For nonzero $\theta$, there is a mismatch compared to the $\theta=0$ case.
This results in an order $\sin^2\theta$ contribution from the triplet $T$, which takes the form
\begin{eqnarray}
\Gamma_{Z\bar\nu_\alpha\nu_\beta}^{(T)}(q^2, M^2) &\simeq& - \Gamma_{Z\bar\nu_\alpha\nu_\beta}^{(1)}(q^2, M^2) - \Gamma_{Z\bar\nu_\alpha\nu_\beta}^{(2)}(q^2, M^2, M^2) - \frac{\partial{\Sigma_{\bar\nu_\alpha\nu_\beta}(p, M^2)}}{\partial{\cancel{p}}} \ .
\end{eqnarray}
The minus sign in front of $\Gamma_{Z\bar\nu_\alpha\nu_\beta}^{(1)}$ comes from $(\cos^2\theta-1)=-\sin^2\theta$, with the $\sin^2\theta$ combines with $y_{\alpha\gamma}y^*_{\beta\gamma}$ to form the coupling factor $\lambda_{\alpha\gamma}\lambda^*_{\beta\gamma}$. The minus sign in front of $\Gamma_{Z\bar\nu_\alpha\nu_\beta}^{(2)}$ comes from $\frac{1}{2}(\cos^4\theta-1)\simeq-\sin^2\theta$ in the small $\theta$ limit. The extra factor $\frac{1}{2}$ is used to eliminate double counting because the two diagrams contributing to $\Gamma_{Z\bar\nu_\alpha\nu_\beta}^{(2)}$ in Eq.~\eqref{eq:diagrams2} are identical after replacing $\phi$ by $T^0$.

Similarly, for the charged-current vertex, we also take into account the above diagrams with $T^0$ replacing $\phi$.
Here, both counterparts of the vertex diagram $\Gamma_{W\bar\ell_\alpha\nu_\beta}$ and self-energy diagram $\Sigma_{\bar\nu_\alpha \nu_\beta}$ are proportional to $y_{\alpha\gamma}y^*_{\beta\gamma}\cos^2\theta$.
In the $\theta=0$ limit, they must be added to the following involving $T^\pm$ and $T^{\pm\pm}$ but not $T^0$,
\begin{align}
&\begin{tikzpicture}
\begin{feynman}
\vertex [label=above:\(W^-\)] (a) at (0,0);
\vertex (b) at (0.7,0);
\vertex (c) at (1.57,0.5);
\vertex (d) at (1.57,-0.5);
\vertex [label=right:\(\ell^-_\alpha\)] (e) at (2.27,0.5);
\vertex [label=right:\(\nu_\beta\)] (f) at (2.27,-0.5);
\feynmandiagram [inline=(a.base)] {
(a)  -- [charged boson] (b),
(b) -- [fermion, edge label=\(\nu^c_\gamma\)] (c),
(d) -- [charged scalar, edge label'=\(T^-\)] (c),
(d) -- [fermion, edge label=\(\ell^+_\gamma\)] (b),
(c) -- [fermion] (e),
(d) -- [anti fermion] (f),
};
\end{feynman}
\end{tikzpicture} 
\hspace{3cm}
\raisebox{0.15cm}{\begin{tikzpicture}
\begin{feynman}
\vertex [label=above:\(W^-\)] (a) at (0,0);
\vertex (b) at (0.7,0);
\vertex (c) at (1.57,0.5);
\vertex (d) at (1.57,-0.5);
\vertex [label=right:\(\ell^-_\alpha\)] (e) at (2.27,0.5);
\vertex [label=right:\(\nu_\beta\)] (f) at (2.27,-0.5);
\feynmandiagram [inline=(a.base)] {
(a)  -- [charged boson] (b),
(b) -- [charged scalar, edge label=\(T^{--}\)] (c),
(d) -- [fermion, edge label'=\(\ell^+_\gamma\)] (c),
(d) -- [charged scalar, edge label=\(T^-\)] (b),
(c) -- [fermion] (e),
(d) -- [anti fermion] (f),
};
\end{feynman} 
\end{tikzpicture}}\\
&\feynmandiagram [inline=(a.base), horizontal=a to d] {
  i1 [particle=\(\nu_\beta\)] -- [fermion] a -- [fermion, edge label'=\(\ell^+_\gamma\)] b -- [fermion] i2 [particle=\(\nu_\alpha\)],
  a -- [charged scalar, half left, looseness=1.5, edge label=\(T^-\)] b,
};
\hspace{1.5cm}
\feynmandiagram [inline=(a.base), horizontal=a to d] {
  i1 [particle=\(\ell^-_\beta\)] -- [fermion] a -- [fermion, edge label'=\(\nu^c_\gamma{,} \ell^+_\gamma\)] b -- [fermion] i2 [particle=\(\ell^-_\alpha\)],
  a -- [charged scalar, half left, looseness=1.5, edge label=\(T^-{,} T^{--}\)] b,
};
\nonumber
\end{align}
in order to render radiative corrections finite and have the correct decoupling behavior.
For nonzero $\theta$ and neglecting the parts suppressed by $q^2/M^2$, the order-$\sin^2\theta$ contribution from the heavy triplet takes the form
\begin{eqnarray}
\Gamma_{W\bar\ell_\alpha\nu_\beta}^{(T)}(q^2, M^2) &\simeq& - \Gamma_{W\bar\ell_\alpha\nu_\beta}(q^2, M^2, M^2) - \frac{1}{2} \frac{\partial{\Sigma_{\bar\nu_\alpha\nu_\beta}(p, M^2)}}{\partial{\cancel{p}}} \ .
\end{eqnarray}

Collecting all the terms, the renormalized gauge couplings are calculated as
\begin{eqnarray}\label{eq:phirenorm2}
    (\delta g_{Z})_{\alpha\beta} &=& g_{Z}^0 \left[ \Gamma_{Z\bar\nu_\alpha\nu_\beta}^{(1)}(q^2, m_\phi^2) + \Gamma_{Z\bar\nu_\alpha\nu_\beta}^{(2)}(q^2, m_\phi^2, M^2) + \Gamma_{Z\bar\nu_\alpha\nu_\beta}^{(T)}(q^2, M^2) +    \frac{\partial{\Sigma_{\bar\nu_\alpha\nu_\beta}(p, m_\phi^2)}}{\partial{\cancel{p}}} \right] \nonumber \\
    &=& g_{Z}^0 \left[ \Gamma_{Z\bar\nu_\alpha\nu_\beta}^{(1)}(q^2, m_\phi^2) + \Gamma_{Z\bar\nu_\alpha\nu_\beta}^{(2)}(q^2, m_\phi^2, M^2) +    \frac{\partial{\Sigma_{\bar\nu_\alpha\nu_\beta}(p, m_\phi^2)}}{\partial{\cancel{p}}} \right. \nonumber\\
    &&\hspace{1.5cm} \left. - \Gamma_{Z\bar\nu_\alpha\nu_\beta}^{(1)}(q^2, M^2) - \Gamma_{Z\bar\nu_\alpha\nu_\beta}^{(2)}(q^2, M^2, M^2) -  \frac{\partial{\Sigma_{\bar\nu_\alpha\nu_\beta}(p, M^2)}}{\partial{\cancel{p}}} \right] \ , \nonumber\\
    (\delta g_{W})_{\alpha\beta} &=& g_{W}^0 \left[ \Gamma_{W\bar\ell_\alpha\nu_\beta}(q^2, m_\phi^2, M^2) + \Gamma_{W\bar\ell_\alpha\nu_\beta}^{(T)}(q^2, m_\phi^2, M^2) + \frac{1}{2} \frac{\partial{\Sigma_{\bar\nu_\alpha\nu_\beta}(p, m_\phi^2)}}{\partial{\cancel{p}}} \right] \nonumber\\
    &=& g_{W}^0 \left[ \Gamma_{W\bar\ell_\alpha\nu_\beta}(q^2, m_\phi^2, M^2) + \frac{1}{2} \frac{\partial{\Sigma_{\bar\nu_\alpha\nu_\beta}(p, m_\phi^2)}}{\partial{\cancel{p}}} \right. \nonumber\\
    &&\hspace{1.5cm} \left. - \Gamma_{W\bar\ell_\alpha\nu_\beta}(q^2, M^2, M^2) - \frac{1}{2} \frac{\partial{\Sigma_{\bar\nu_\alpha\nu_\beta}(p, M^2)}}{\partial{\cancel{p}}} \right] \ .
\end{eqnarray}
It is obvious that these results are UV finite and have the correct decoupling behavior when $m_\phi = M \to\infty$.
It is worth noting that both $\delta g_{Z}$ and $\delta g_{W}$ are equal to the above $\phi$ diagram contributions subtracting the same set of diagrams with $\phi$ replaced by $T^0$. Their structure is consistent with the argument made in the appendix of~\cite{Zhang:2024meg}.

Relevant for phenomenological studies, we present the renormalized gauge couplings in the heavy triplet limit while keeping $\phi$ light, $M^2\gg m_\phi^2, |q^2|$,
\begin{align}
&\hspace{-0.3cm}(\delta g_Z)_{\alpha\beta}
= - \frac{g_Z^0\lambda_{\alpha\gamma}\lambda^*_{\beta\gamma}}{32\pi^2} \label{eq:6}\\
&\quad\times\left\{ \ln\frac{M^2}{m_\phi^2} - \frac{11}{2} + 2\int dx dy \left[\ln \left(\frac{M^2}{(1-x-y)m_\phi^2-xy q^2}\right)  + \frac{x y q^2}{(1-x-y)m_\phi^2-xy q^2} 
\right]\right\} \ \nonumber,
\end{align}
and
\begin{align}\label{eq:7}
(\delta g_W)_{\alpha\beta} = - \frac{g_W^0 \lambda_{\alpha\gamma}\lambda^*_{\beta\gamma}}{64\pi^2} \left(\ln\frac{M^2}{m_\phi^2} - 2 \right) \ .
\end{align}

An interesting observation of Eqs.~\eqref{eq:6} and \eqref{eq:7} is that the renormalized neutrino-$Z$-boson coupling $(g_Z)_{\alpha\beta}$ depends on the momentum transfer $q^2$ but $(g_W)_{\alpha\beta}$ does not.
Such a dependence arises due to the separation between $m_\phi$ and $M$ mass scales.
It is useful to examine the limiting behaviors of $(g_Z)_{\alpha\beta}$ as a function of $Q \equiv \sqrt{-q^2}$ which is the momentum transfer in neutrino neutral-current scattering processes,
\begin{align}
\label{eq:asumptotic}
    \frac{(g_Z)_{\alpha\beta}}{g_Z^0} = \delta_{\alpha\beta} - \frac{\lambda_{\alpha\gamma}\lambda^*_{\beta\gamma}}{32\pi^2}\times \left\{ \begin{array}{ll}
    \ln \frac{M^2}{Q^2} + \ln \frac{M^2}{m_\phi^2} - \frac{7}{2} \ , &\hspace{0.3cm} Q \gg m_\phi \\
    2\ln \frac{M^2}{m_\phi^2} -4 \ , &\hspace{0.3cm} Q \ll m_\phi
    \end{array} \right.\quad .
\end{align}
If $\phi$ is the only beyond-SM particle with mass well below the EW scale, $g_Z$ runs logarithmically with $Q$ and gets smaller at lower energies. This running effect ceases after $Q$ falls below the $m_\phi$ threshold.

\begin{figure}[t]
    \centering
    \includegraphics[width=0.618\linewidth]{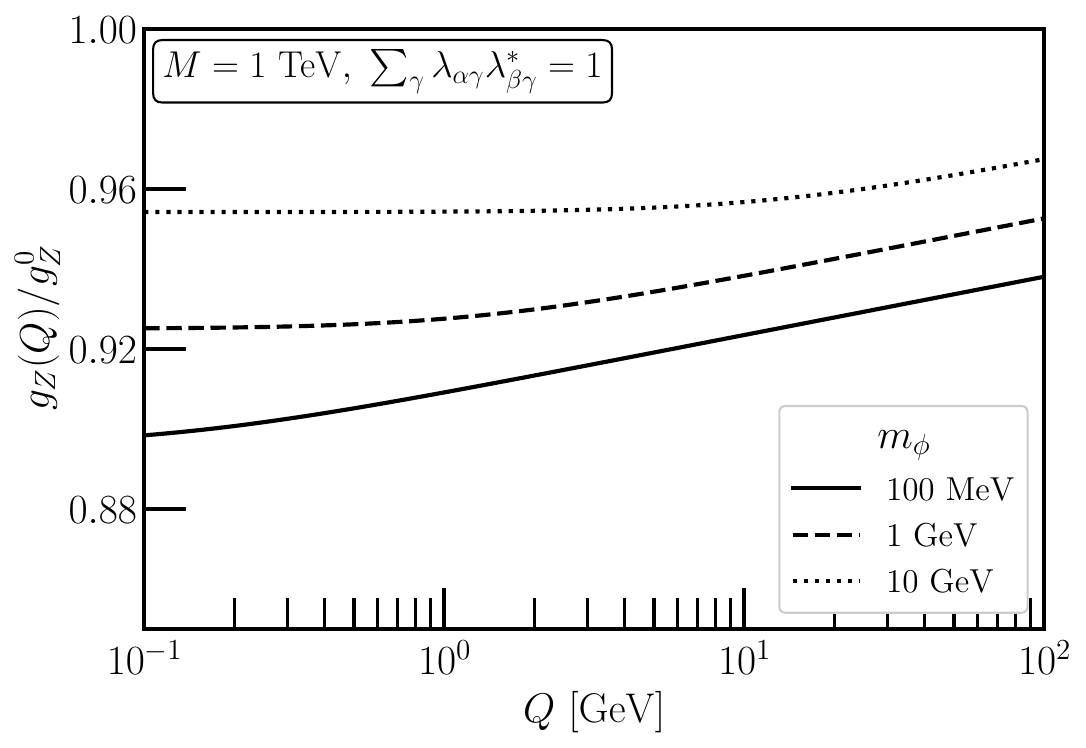}
    \caption{Radiative correction to the $Z\nu\bar\nu$ coupling as a function of momentum transfer $Q$ in the presence of a light scalar $\phi$ that mediates neutrino self-interaction. $M=1\,$TeV, $\lambda_{\alpha\gamma}\lambda_{\beta\gamma}^*=1$.}\label{fig:1}
\end{figure}
\cref{fig:1} depicts the energy-scale dependence of $g_{Z}^2$ for three choices of $m_\phi$ and with the coupling $\lambda$ set to unity.
Here for simplicity, we set the gauge coupling parameters at the $Z$-pole to their best fit values in the Standard Model~\footnote{A more careful analysis including global fit to all EW observables in the singlet-triplet model for neutrino self-interaction will be presented in a separate work.}, so that the couplings of $Z$ boson to fermions other than neutrinos remain intact.
In contrast, due to the radiative correction from $\phi$, the $Z\nu\bar\nu$ coupling deviates from the Standard Model prediction even at the $Z$-pole. Such a deviation can be constrained by the invisible $Z$-boson decay width measurement by LEP-II, as will be discussed in~\cref{sec:Z}.

\section{Implications for Weak Mixing Angle Measurements}\label{sec:WMA}

Measuring the weak mixing angle $\sin^2\theta_W$ and its running with respect to momentum-transfer scales is one of the most powerful tests of the EW theory.
So far, the most precise measurements come from electron-electron and electron-hadron scattering processes~\cite{Kumar:2013yoa}.
NuTeV measured this quantity using neutrino scattering, yielding a higher value of $\sin^2\theta_W$ than expected for GeV-scale momentum transfer~\cite{NuTeV:2001whx,Kumar:2013yoa}.
As we will show in this section, the neutrinophilic scalar $\phi$ can impact neutrino-scattering observables in nontrivial ways, predicting deviations of inferred measurements of $\sin^2\theta_W$ depending on the type of probe.
We will explore two specific probes quantitatively -- neutrino Deep-Inelastic Scattering ($\nu$DIS) and coherent elastic neutrino-nucleus scattering (CEvNS) -- and provide some discussion surrounding a third, electron-neutrino elastic scattering (e$\nu$ES).

Because most neutrino-centric measurements of $\sin^2\theta_W$ are probed using muon-neutrino beams, we will focus on $\nu_\mu$-philic self-interactions for the remainder of this section. We will assume that the interaction of $\phi$ takes the following structure,
\begin{equation}
    \lambda_{\alpha\beta} = \lambda \delta_{\alpha \mu} \delta_{\beta\mu} \ .
\end{equation}

\begin{figure}[!htbp]
    \centerline{
\includegraphics[width=\linewidth]{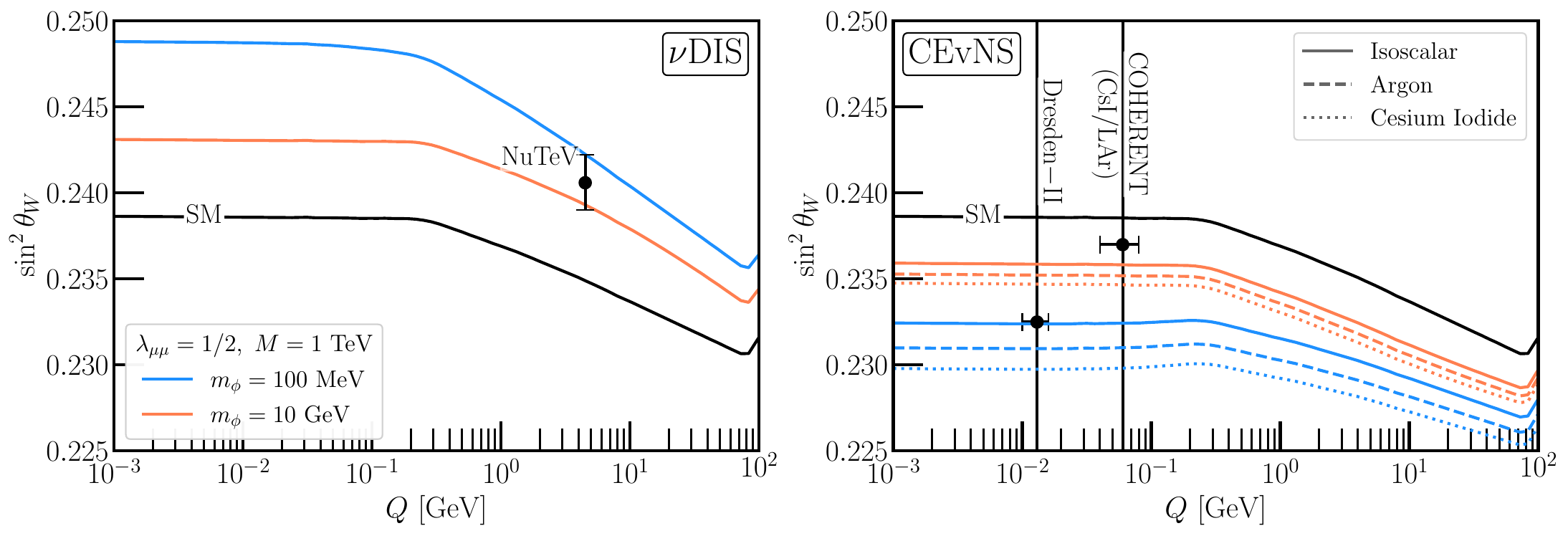}}
    \caption{If one absorbs the novel $Q$-dependence of $g_Z$ into the running of the weak mixing angle, the corresponding values of $\sin^2\theta_W$ needed for interpret the neutrino DIS and CE$\nu$NS experiments are shown by the left and right panels, respectively, taking $M=1\,$TeV, $\lambda_{\mu\mu}=0.5$. The left-panel includes the measurement using $\nu$DIS from NuTeV~\cite{NuTeV:2001whx}; the right includes measurements determined from the COHERENT~\cite{DeRomeri:2022twg} and Dresden-II~\cite{AristizabalSierra:2022axl} experiments' results.}
    \label{fig:2}
\end{figure}

\textbf{Neutrino Deep-Inelastic Scattering ($\mathbf{\nu}$DIS):} For high-energy neutrino scattering (still with center-of-mass energy well below the EW scale), one can infer the running of the weak mixing angle by measuring both neutral- and charged-current neutrino cross sections for scattering on isoscalar targets~\cite{NuTeV:1999kup, NuTeV:2001whx}.
%For $\nu$-DIS, we consider NuTeV-like experiemnts with center of mass energy well below the electroweak scale. In this case, one can define the ratio of neutral current to charged current cross sections for neutrino or anti-neutrino scattering on isoscalar targets~\cite{NuTeV:1999kup, NuTeV:2001whx}.
An often-adopted observable is the Paschos-Wolfenstein parameter, which is directly related to $\sin^2\theta_W$ at the corresponding energy scale~\cite{Paschos:1972kj}
\begin{equation}
    \begin{split}
        R^- \equiv \frac{\sigma_{\nu_\mu N \to \nu_\mu X} - \sigma_{\bar \nu_\mu N \to \bar\nu_\mu X} }{\sigma_{\nu_\mu N \to \mu^- X} - \sigma_{\bar\nu_\mu N \to \mu^+ X}} \simeq \frac{1}{2} - \sin^2\theta_W(Q) \ .
    \end{split}
\end{equation}
%This relation holds at tree-level and assuming the target is comprised of only first generation of quarks.
In the presence of neutrino self-interaction, the neutral and charged current interaction cross sections are rescaled by the ratios $(g_Z(Q)/g_Z^0)^2$ and $(g_W/g_W^0)^2$, respectively. In turn, the predicted $R^-$ value is modified. If one insists to attribute such an effect to the running of the weak mixing angle, they would interpret the result as
\begin{equation}
     (\sin^2\theta_W)_{\rm apparent} = \frac{1}{2} - \left( \frac{1}{2} - \sin^2\theta_W(Q) \right) \frac{(g_Z(Q)/g_Z^0)^2}{(g_W/g_W^0)^2} \ .
\end{equation}
On the right-hand side, $\sin^2\theta_W(Q)$ stands for the actual weak mixing, which is equal to $(\sin^2\theta_W)_{\rm apparent}$ in the absence of neutrino self-interaction. Because the radiative correction in Eqs.~\eqref{eq:6} reduces $g_Z$ at lower $Q$, the resulting
$(\sin^2\theta_W)_{\rm apparent}$ is larger than the actual value.
This effect is shown by the left panel of Fig.~\ref{fig:2}.
The black data point shows the measurement from NuTeV~\cite{NuTeV:2001whx} -- if the higher-than-expected measurement is interpreted as evidence for self-interactions, it points to couplings $\lambda_{\mu\mu} \approx 0.5$ for mediator masses $m_\phi \approx 10$~GeV.

\textbf{Coherent Elastic Neutrino-Nucleus Scattering (CEvNS):}
At much lower momentum-transfer scales, neutrinos can interact coherently with the entire nucleus in the process now referred to as `CEvNS.'
This scattering process, first measured in 2017~\cite{COHERENT:2017ipa}, provides a complementary neutrinophilic test of the running of $\sin^2\theta_W$ at much lower $Q$.

The differential cross section for CEvNS as a function of the nucleus's recoiling kinetic energy $T$ is
\begin{equation}
    \frac{d \sigma_{\nu N\to \nu N}}{dT} = \frac{G_F^2 Q_W^2 M}{8\pi} \left(2-\frac{M T}{E_\nu^2}\right)F(Q)^2 \ ,
\end{equation}
where $E_\nu$ is the incoming neutrino energy, $M$ is the target nucleus mass, and $F$ is the nuclear form factor. 
The weak charge is $Q_W=N-Z(1-4\sin^2\theta_W(Q))$ where $N, Z$ are the number of neutrons and protons inside the nucleus.
Introducing the neutrino self-interactions affects the neutrino-$Z$ boson coupling and the cross section is further rescaled by a factor of $(g_Z(Q)/g_Z^0)^2$.
Again, if one insists to attribute it to the running of the weak mixing angle, they would interpret the result as
\begin{equation}
     (\sin^2\theta_W)_{\rm apparent} = \frac{1}{4} \left[ 1 - \frac{N}{Z} + \frac{g_Z(Q)}{g_Z^0} \left( 4 \sin^2\theta_W(Q) -1 + \frac{N}{Z} \right) \right]   \ .
\end{equation}
Interestingly, the radiative correction from Eqs.~\eqref{eq:6} that reduces $g_Z(Q)$ at low energies is now interpreted as a smaller $(\sin^2\theta_W)_{\rm apparent}$ than the actual value, which is the opposite to the finding with the $\nu$-DIS experiment. We show this behavior for a few choices of $(N,Z)$ in the right panel of~\cref{fig:2}.

By performing both the $\nu$-DIS and CE$\nu$NS experiments precisely, one can break the degeneracy between the usual scale dependence of $\sin^2\theta_W$ and the novel radiative correction to $Z\nu\bar\nu$ coupling from neutrino self-interaction.

\textbf{Electron-neutrino Elastic Scattering (e$\nu$ES):} Additionally, the weak mixing angle can be extracted by measuring neutrino-electron scattering in a neutrino-beam environment~\cite{deGouvea:2019wav,Alves:2024twb,deGouvea:2025zfq}.
These measurements exploit the electron-recoil-energy dependence of this scattering process, as well as differences between (for instance) $\nu_\mu$ and $\nu_e$ differential cross sections, to perform such a measurement.
Reinterpreting such a result in the context of adjusted $g_Z(Q)$ is not straightforward -- a dedicated analysis of e$\nu$ES is necessary to determine the sensitivity of this probe to the effect of neutrino self-interactions and we leave it to future, dedicated work.
Nevertheless, given the momentum transfers probed by e$\nu$ES and the projected sensitivities presented in Refs.~\cite{deGouvea:2019wav,Alves:2024twb,deGouvea:2025zfq}, we expect that this probe will be comparable to, and even more powerful than, CEvNS measurements in this context.

\section{Constraint from Invisible $Z$ Width Measurement}\label{sec:Z}

We revisit the constraint on neutrino self-interaction from the LEP-II measurement of invisible $Z$ boson decay width~\cite{ALEPH:2005ab}.
For light $\phi$, this constraint was first estimated based on the tree-level decay channel $Z\to \nu\nu\phi$~\cite{Berryman:2018ogk, Lyu:2020lps}.
Subsequent references attempted to include loop corrections to $Z\to\nu\bar\nu$ decay. However, we find discrepancies between our results and the past literature.
Here, we present our calculation of the invisible $Z$ decay width in the context of the UV complete model presented in~\cref{sec:model}.

We first present the radiative correction to $Z\to\nu_\alpha\bar\nu_\beta$ decay, using the result of $\delta g_Z$ given~\cref{eq:6}.
Starting from the tree-level $Z\to\nu\bar\nu$ decay width, $\Gamma^{\rm tree}_{Z\to\nu\bar\nu}=(g_Z^0)^2M_Z/(24\pi)$, the correction due to $\delta g_Z$ is equal to
\begin{equation} 
\begin{split}
    \delta \Gamma_{Z\to\nu_\alpha\bar\nu_\beta} &= \Gamma^{\rm tree}_{Z\to\nu\bar\nu} \frac{2{\rm Re}(\delta g_Z)_{\alpha\beta}}{g_Z^0} = - \frac{\sqrt{2} G_F M_Z^3 \lambda_{\alpha\gamma}\lambda_{\beta\gamma}^*}{384\pi^3} \left\lbrace \ln\frac{M^2}{m_\phi^2} - \frac{11}{2}\right. \\
    &\left. + 2 \mathrm{Re}\int_{0}^{1} dx \int_{0}^{1-x} dy \left[ \ln\left(\frac{M^2}{(1-x-y)m_\phi^2 - xyM_Z^2}\right) + \frac{xyM_Z^2}{(1-x-y)m_\phi^2 - xyM_Z^2} \right]  \right\rbrace \ , 
\end{split}
\end{equation}
where we used $g_Z^0=(\sqrt2 G_F M_Z^2)^{1/2}$.
If the final-state flavors $\alpha\neq\beta$, one also needs to add the above result to the decay width of the charge-conjugate channel $Z\to \nu_\beta \bar\nu_\alpha$.
Importantly, our result is free from UV divergences. In contrast, a previous work~\cite{Dev:2024ygx} did not work with a gauge-invariant model and lacked explicit UV-divergence cancellation, leaving the finite part of the radiative correction ambiguous.

In addition to the loop process, a light $\phi$ can also be radiated by the final state neutrino or anti-neutrino, leading to three-body $Z$-boson decay $Z\to\nu_\alpha\nu_\beta\phi^*$ that occurs at the same $\mathcal{O}(\lambda^2)$ order.
It is useful to provide some details of our calculation.
For a given momentum assignment, 
there are always two diagrams contributing,
\begin{equation}\label{eq:S3}
\begin{split}
\begin{tikzpicture}
\begin{feynman}
\vertex [label=above:\(Z\)] (a) at (0,0);
\vertex (b) at (0.7,0);
\vertex (c) at (1.3,0.32);
\vertex [label=right:\(\nu_\alpha\)] (d) at (2,0.67);
\vertex [label=right:\(\phi\)] (e) at (2,0);
\vertex [label=right:\(\bar\nu^c_\beta\)] (f) at (1.5,-0.5);
\feynmandiagram [inline=(a.base)] {
(a)  -- [photon] (b),
(b) -- [fermion, edge label=\(\nu^c_\beta\)] (c),
(c) -- [fermion] (d),
(e) -- [charged scalar] (c),
(f) -- [fermion] (b),
};
\end{feynman}
\end{tikzpicture}\hspace{1cm} \raisebox{0.65cm}{+}\hspace{1cm}
\begin{tikzpicture}
\begin{feynman}
\vertex [label=above:\(Z\)] (a) at (0,0);
\vertex (b) at (0.7,0);
\vertex (c) at (1.3,-0.32);
\vertex [label=right:\(\bar\nu^c_\beta\)] (d) at (2,-0.67);
\vertex [label=right:\(\phi\)] (e) at (2,0);
\vertex [label=right:\(\nu_\alpha\)] (f) at (1.5,0.5);
\feynmandiagram [inline=(a.base)] {
(a)  -- [photon] (b),
(c) -- [fermion, edge label=\(\nu_\alpha\)] (b),
(d) -- [fermion] (c),
(e) -- [charged scalar] (c),
(b) -- [fermion] (f),
};
\end{feynman}
\end{tikzpicture}
\end{split}
\end{equation}
The corresponding decay amplitude for $Z(q) \to \phi^*(p_1)+\nu_\alpha(p_2)+\nu_\beta(p_3)$ reads
 \begin{equation} \label{eq:S4}
   i\mathcal{M} = \frac{i \lambda_{\alpha\beta} g_Z^0}{(p_1+p_2)^2}\bar u(p_2) (\cancel{p}_1+\cancel{p}_2)\gamma_\mu \mathbb{P}_R v(p_3) \varepsilon_Z^\mu(q) 
   - \frac{i \lambda_{\alpha\beta} g_Z^0}{(p_1+p_3)^2} \bar u(p_2) \gamma_\mu \mathbb{P}_L (-\cancel{p}_1-\cancel{p}_3) v(p_3) \varepsilon_Z^\mu(q) \ ,
\end{equation}
where the relative minus sign between the two terms arises because the $Z\nu\bar\nu$ and $Z\nu^c\bar\nu^c$ couplings are opposite.
When writing down the above decay amplitudes, we treat the final-state $\nu_\beta$ as the antiparticle of $\nu_\beta^c$ field.
The resulting three-body decay rate (adding the charge-conjugate channel $Z\to \bar\nu_\alpha\bar\nu_\beta \phi^*$) is (for $\alpha\neq \beta$)
\begin{align}
    \Gamma_{Z\to\nu_\alpha\nu_\beta \phi^*} + \Gamma_{Z\to\bar\nu_\alpha\bar\nu_\beta \phi} &= \frac{\sqrt{2} G_F M_Z^3 |\lambda_{\alpha\beta}|^2}{192\pi^3} \left\lbrace \left[\ln{\left(\frac{M_Z^2}{m_\phi^2}\right)} - \frac{11}{3}\right] + \frac{8m_\phi^2}{M_Z^2}\left[\ln{\left(\frac{M_Z^2}{m_\phi^2}\right)} - \frac{5}{8}\right] \right. \nonumber\\
&\hspace{3cm} + \frac{2m_\phi^4}{M_Z^4} \left[\ln{\left(\frac{M_Z^2}{m_\phi^2}\right)} \left(
\ln{\left(\frac{M_Z m_\phi}{M_Z^2+m_\phi^2}\right)} + \frac{3}{2}\right) + \frac{9}{2} \right. \nonumber\\
&\left.\left.  \hspace{3cm} 
+{\rm Li}_2\left( \frac{M_Z^2}{M_Z^2+m_\phi^2}\right)
-{\rm Li}_2\left( \frac{m_\phi^2}{M_Z^2+m_\phi^2}
\right)
\right] - \frac{m_\phi^6}{3M_Z^6}\right\rbrace \ .
\end{align}
Note, if the final state neutrino flavors are identical $\alpha=\beta$, the above three-body decay rates should be multiplied by an extra factor of 1/2. Our result differs from~\cite{Brdar:2020nbj, Dev:2024ygx}. 
%whose result would be obtained if only one of the decay diagrams in Eq.~\eqref{eq:S3} was taken into account.

The PDG-averaged $Z$-boson invisible decay width is $499.3\pm1.5$\,MeV~\cite{ParticleDataGroup:2024cfk}, which leads to the upper bound
\begin{equation}
    \sum_{\alpha\beta}\left[(2-\delta_{\alpha\beta}) \delta \Gamma_{Z\to\nu_\alpha\bar\nu_\beta}  + \frac{\Gamma_{Z\to\nu_\alpha\nu_\beta \phi^*} + \Gamma_{Z\to\bar\nu_\alpha\bar\nu_\beta \phi}}{1+\delta_{\alpha\beta}}\right] < 3\,{\rm MeV} \ ,
\end{equation}
at 2$\sigma$ CL. Both terms on the left-hand side occur at $\mathcal{O}(\lambda^2)$ order.
%The upper limit on $|\lambda|$ from invisible $Z$ width are shown in Fig.~\ref{fig:4}.

\begin{figure}[t]
\includegraphics[width=0.49
\linewidth]{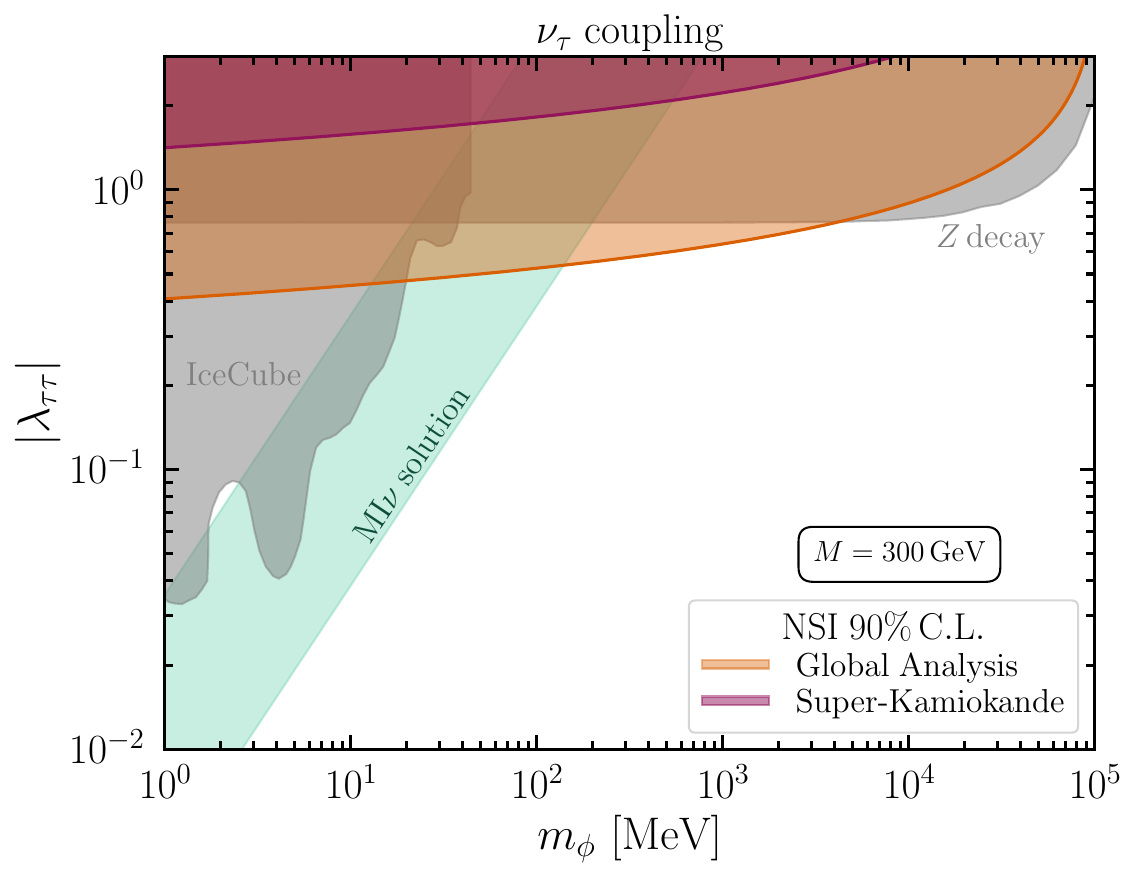} \includegraphics[width=0.49\linewidth]{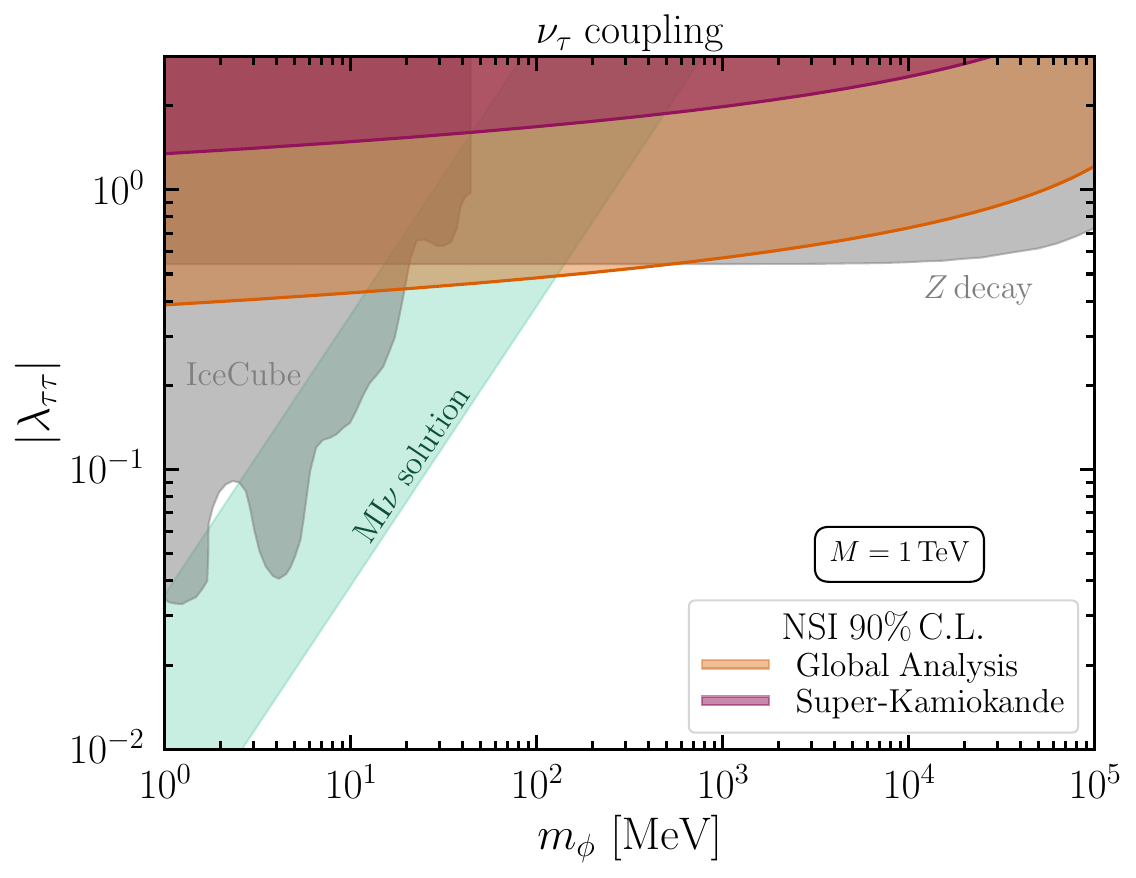}
\caption{Sensitivity to $\nu_\tau$ self-interactions from existing experimental bounds for two values of the scalar triplet mediator mass (left and right panels). Shaded regions show limits derived from a global analysis of non-standard neutrino interactions using oscillation and scattering data~\cite{Coloma:2023ixt} and from atmospheric neutrino data in Super-Kamiokande~\cite{Super-Kamiokande:2011dam}, both at $90\%$ CL. Current bounds at the $2\sigma$ C.L. from other sources are shown in gray, including IceCube measurements of ultra-high-energy astrophysical neutrinos~\cite{Esteban:2021tub}, the invisible $Z$ decay width (revisited in this work), and cosmologically relevant regions corresponding to the moderately interacting neutrinos (MI$\nu$) solution, indicated by the light green band.
}\label{fig:4}
\end{figure}

To present constraint, need to make choices of $\lambda$ flavor structure. We consider two cases in this work: 
1) $\nu_\tau$-philic self-interaction
\begin{equation}
    \sum_{\alpha,\beta,\gamma}\lambda_{\alpha\gamma}\lambda_{\beta\gamma}^* = \sum_{\alpha,\beta}|\lambda_{\alpha\beta}|^2 = |\lambda_{\tau\tau}|^2
\end{equation}
2) flavor universal self-interaction
\begin{equation}
    \sum_{\alpha,\beta,\gamma}\lambda_{\alpha\gamma}\lambda_{\beta\gamma}^* = \sum_{\alpha,\beta}|\lambda_{\alpha\beta}|^2 = |\lambda_{ee}|^2 + |\lambda_{\mu\mu}|^2 + |\lambda_{\tau\tau}|^2 \equiv |\lambda|^2
\end{equation}
The corresponding $Z$-decay constraints at $2\sigma$~CL are shown in Figs.~\ref{fig:4} and \ref{fig:borexino_2sigCL}, respectively.

\section{Constraint from Non-Standard Neutrino Interactions}\label{sec:NSI}

In this section, we explore a $\lambda$ flavor structure of interest to cosmology where the self-interaction only applies for tau-neutrinos,
\begin{equation}\label{eq:nutauphilic}
\lambda_{\alpha\beta} = \lambda \delta_{\alpha\tau} \delta_{\beta\tau} \ .
\end{equation}
Due to the violation of flavor universality, we identify a relevant constraint on this scenario from experiments probing non-standard interactions (NSI) between neutrinos and matter.
The neutrino self-interaction affects the $Z$-mediated neutral currents at one-loop level. 

The neutrino-matter interaction potential, allowing for NSIs, is commonly parameterized as
\begin{equation}
V_{\alpha\beta} = \sqrt{2} G_F n_e 
\begin{pmatrix}
1 + \epsilon_{ee} & \epsilon_{e\mu} & \epsilon_{e\tau}\\
\epsilon_{\mu e} & \epsilon_{\mu\mu} & \epsilon_{\mu\tau}\\
\epsilon_{\tau e} & \epsilon_{\tau\mu} & \epsilon_{\tau\tau}
\end{pmatrix} \ .
\end{equation}
The standard Mikheyev–Smirnov–Wolfenstein potential~\cite{Wolfenstein:1977ue, Mikheyev:1985zog} corresponds to the case of all $\epsilon_{\alpha\beta}=0$.
The $\nu_\tau$-philic self-interaction gives radiative correction to the $Z\nu_\tau \bar\nu_\tau$ coupling, $(\delta g_Z)_{\tau\tau}$, given by~\cref{eq:6}.
In turn, the only nonzero NSI parameter $\epsilon_{\tau\tau}$ originates from the deviation of $\nu_\tau$-matter interaction via the $Z$-boson exchange from the SM prediction,
\begin{equation}\label{eq:Vtautau}
V_{\tau\tau} \simeq \frac{(\delta g_Z)_{\tau\tau} g_Z^0}{M_Z^2}  \left( g_V^e n_e + g_V^u n_u + g_V^d n_d \right) \ ,
\end{equation}
where $g_Z^0 = g/(2\cos\theta_W)$, $G_F = (g_Z^0)^2/(\sqrt{2} M_Z^2)$, and the vector couplings are
\begin{equation}
g_V^e = -\frac{1}{2}+2\sin^2\theta_W\ , \quad g_V^u=\frac{1}{2}-\frac{4}{3}\sin^2\theta_W\ , \quad g_V^d = -\frac{1}{2} + \frac{2}{3}\sin^2\theta_W \ .
\end{equation}
For simplicity, we assume isoscalar matter with an equal number of proton and neutrons, thus $n_u=n_d=3n_e$.
As a result, the bracket in~\cref{eq:Vtautau} is equal to $(-1/2)$ and we obtain
\begin{equation}
\epsilon_{\tau\tau} = - \frac{(\delta g_Z)_{\tau\tau}}{2g_Z^0} = \frac{|\lambda|^2}{32\pi^2}
\left(\ln \frac{M^2}{m_\phi^2} - 2\right) \ .
\end{equation}

The Super-Kamiokande experiment has set an upper bound on the NSI~\cite{Super-Kamiokande:2011dam, Ohlsson:2012kf}
\begin{equation}
|\varepsilon_{\mu\mu}-\varepsilon_{\tau\tau}| < 0.049 \times 3 \ .
\end{equation}
A more recent analysis that performs a global fit to neutrino oscillation measurements obtained a much stronger constraint~\cite{Coloma:2023ixt}
\begin{equation}
-0.0215<\varepsilon_{\tau\tau}- \varepsilon_{\mu\mu} < 0.0122 \ .
\end{equation}
Both limits are set at 90\% CL. We show the corresponding constraint on the $\lambda_{\tau\tau}$ versus $m_\phi$ parameter space in Fig.~\ref{fig:4}. The results have logarithmic dependence on the heavy scale $M$. Remarkably, for $m_\phi$ below ${\sim}$GeV scale, the neutrino NSI constraint can be stronger than that from the invisible $Z$ decay width and gives the leading constraint on the neutrino self-interaction strength.
It is also complementary to the IceCube probe using ultra-high energy astrophysical neutrinos which sets the leading constraint for $m_\phi \lesssim 20\,$MeV~\cite{Esteban:2021tub}.

\section{Constraint from Solar Neutrino Flux Measurement}\label{sec:solar}
In this section, we explore the solar-neutrino flux measurement recently made by the Borexino experiment using elastic scattering off electrons in the detector.
The high-precision spectroscopy of solar neutrinos achieved by Borexino~\cite{BOREXINO:2018ohr}, combined with global flux determinations~\cite{BOREXINO:2022abl} and comparisons with Standard Solar Models (SSMs)~\cite{Gonzalez-Garcia:2023kva} makes Borexino a sensitive probe of new physics in the neutrino sector~\cite{Coloma:2022umy,Kelly:2024tvh}.
In this section, we examine its potential to constrain neutrino self-interactions (\(\nu\)SI).

For clarity of the discussion, we consider flavor-diagonal/universal couplings for the neutrino self-interaction,
\begin{equation}
\lambda_{\alpha\beta} = \lambda \delta_{\alpha\beta} \ .
\end{equation}
In this case, the radiative correction to neutrino gauge couplings,~\cref{eq:6,eq:7}, are flavor-diagonal and their coefficients simplify to $\lambda_{\alpha\gamma}\lambda^*_{\beta\gamma} = |\lambda|^2$.
Moreover, because the neutrinophilic scalar $\phi$ mass is constrained to be heavier than several MeV by big-bang nucleosynthesis, the solar-neutrino/electron scattering occurs with $Q\ll m_\phi$. The renormalized gauge couplings are well approximated by
\begin{equation}\label{eq:gratios}
\frac{g_Z}{g_Z^0} = 1 - \frac{|\lambda|^2}{16\pi^2}
\left(\ln \frac{M^2}{m_\phi^2} -2\right) , \quad\quad
\frac{g_W}{g_W^0} = 1 - \frac{|\lambda|^2}{64\pi^2} \left(\ln\frac{M^2}{m_\phi^2} - 2 \right) \ .
\end{equation}
In our numerical analysis, we use the more general formula~\cref{eq:6} for $g_Z$.

For MeV-scale solar neutrinos detected by Borexino, it is a good approximation to assume the target electron to be at rest and ignore its atomic binding energy.
The differential scattering cross section for $\nu_\alpha e^- \to \nu_\alpha e^-$ is
\begin{equation}\label{eq:EESXS}
\frac{\mathrm{~d}\sigma_{\nu_\alpha e}}{\mathrm{~d}T}(E_\nu, T) =
\frac{2 G_F^2 m_e}{\pi}
\Bigg[
(g_{L}^{\alpha})^2
+ (g_{R}^{\alpha})^2 \left(1 - \frac{T}{E_\nu}\right)^2  - (g_{L}^{\alpha})(g_{R}^{\alpha})\,\frac{m_e T}{E_\nu^2} 
\Bigg] \ ,
\end{equation}
where $T$ is the final-state electron recoil energy. For a given incoming neutrino energy $E_\nu$, the range of $T$ is
\begin{equation}
0\leq T \leq \frac{2 E_\nu^2}{m_e + 2 E_\nu} \ . 
\end{equation}
For given recoil $T$, the smallest incoming neutrino energy for the scattering to happen is
\begin{equation}
E_\nu^{\rm min}(T) = \frac{T +\sqrt{T^2 + 2m_eT}}{2} \ .
\end{equation}

The couplings $g_L^\alpha$ and $g_R^\alpha$ depend on the neutrino flavor $\alpha = e, \mu, \tau$ and are affected in the presence of neutrino self-interaction,
\begin{equation}
g_L^\alpha = \frac{g_Z}{g_Z^0} \left(\sin^2\theta_W - \frac{1}{2} \right) + \frac{g_W}{g_W^0}\delta_{\alpha e}, \quad
g_R^\alpha = \frac{g_Z}{g_Z^0} \sin^2\theta_W \ ,
\end{equation}
where the ratios $g_Z/g_Z^0$ and $g_W/g_W^0$ are given in~\cref{eq:gratios}. 

\begin{figure}[t]
\centering
\includegraphics[width=0.8\textwidth]{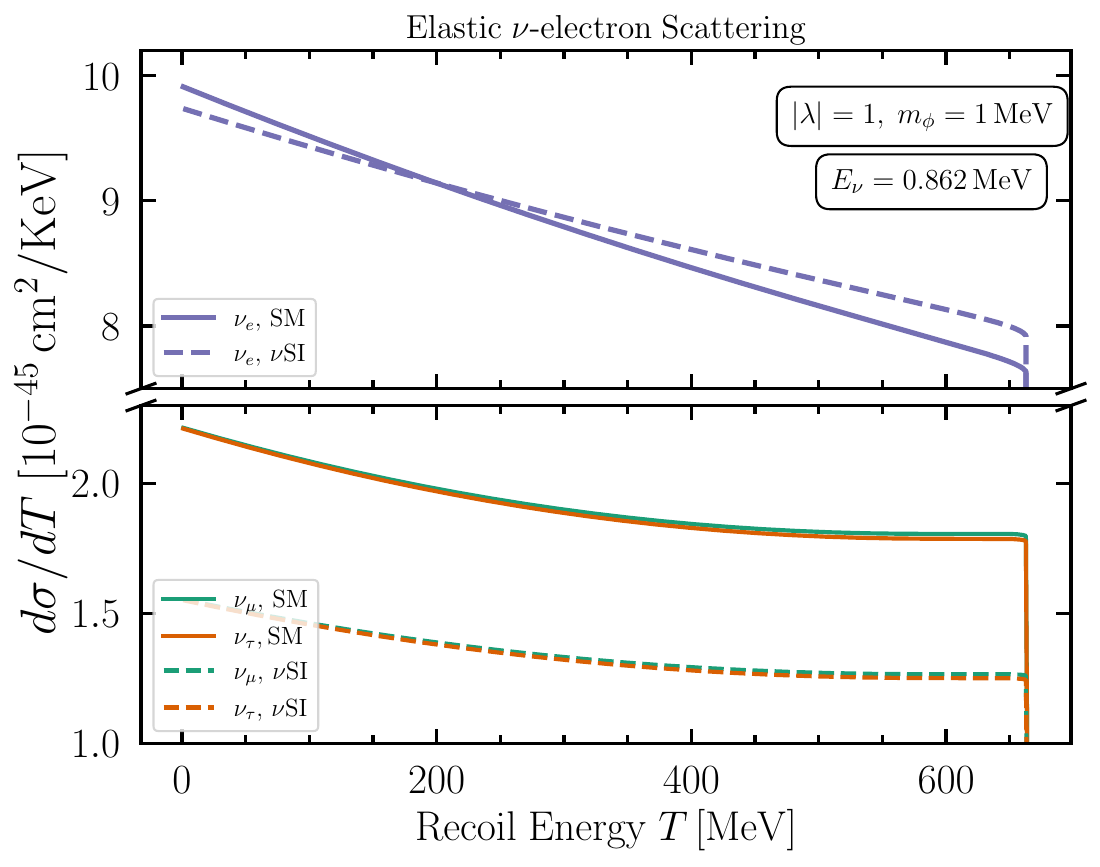}
\caption{Differential cross sections for neutrino–electron elastic scattering, including electroweak radiative corrections (solid) and in the presence of $\nu$SI (dashed), shown for each neutrino flavor. The effect of $\nu$SI reduces the cross section for muon and tau neutrinos and induces a slight tilt in the electron-neutrino spectrum due to the charged-current contribution.}
\label{fig:Diff_XS_elastic}
\end{figure}

SM, EW radiative corrections to the neutrino-electron elastic scattering cross section~\cite{Tomalak:2019ibg} are included in our analysis.
The corrections modify the tree-level cross sections by including virtual- and soft-photon effects, bremsstrahlung, QED vertex contributions, non-factorizable terms, and dynamical corrections from fermion loops. 
Building on these radiative corrections, we next incorporate the modification to the differential cross sections arising from a neutrinophilic mediator, as given by~\cref{eq:gratios}.
In~\cref{fig:Diff_XS_elastic}, we show the differential cross sections for $\nu_e$, $\nu_\mu$, and $\nu_\tau$ electron elastic scattering, comparing the SM predictions (with EW radiative corrections) to those in the presence of $\nu$SI, which results in a reduction of the cross section for $\nu_\mu$ and $\nu_\tau$ and a $T$-dependent modification of the electron-neutrino recoil spectrum.
For electron neutrinos, since both neutral-current and charged-current interactions contribute, in the presence of $\nu$SI, the coupling $g_L^e$ remains approximately unchanged, while $g_R^e$ decreases.
At low recoil energies, where the terms in the elastic scattering cross section are comparable, this leads to a reduction in the combination of the first two terms in~\cref{eq:EESXS} and therefore a decrease in the differential cross section near the low-$T$ endpoint.
At high recoil energies ($T \simeq E_\nu$), the combination of non-vanishing terms increases slightly, resulting in a tilt in the recoil spectrum for electron neutrinos.
For muon and tau neutrinos interacting only via neutral currents, the decrease in left- and right-handed couplings is the same, directly reducing the cross section across the spectrum.   

We analyze Borexino Phase-III data following Refs.~\cite{Coloma:2022umy,Kelly:2024tvh}.
The dataset includes the monoenergetic lines from \( ^7\mathrm{Be} \) (\(E_\nu{=}0.862~\rm{MeV}\)) and pep (\(E_\nu{=}1.44~\rm{MeV}\)) neutrinos, as well as the continuous CNO spectrum up to \(1.74~\rm{MeV}\).
Borexino, a 278-ton ultra-pure liquid scintillator detector at LNGS, measures solar neutrinos via elastic scattering on electrons.
Its spectral analysis distinguishes contributions from \(\mathrm{pp}, \mathrm{pep}, \mathrm{CNO}, ^8\mathrm{B},\) and \( ^7\mathrm{Be} \) fluxes.
Among these, the \( ^7\mathrm{Be} \) flux is most precisely measured, with a 2.7\% uncertainty reported in Phase-II.
\begin{figure}[t]
\centering
\includegraphics[width=0.8\textwidth]{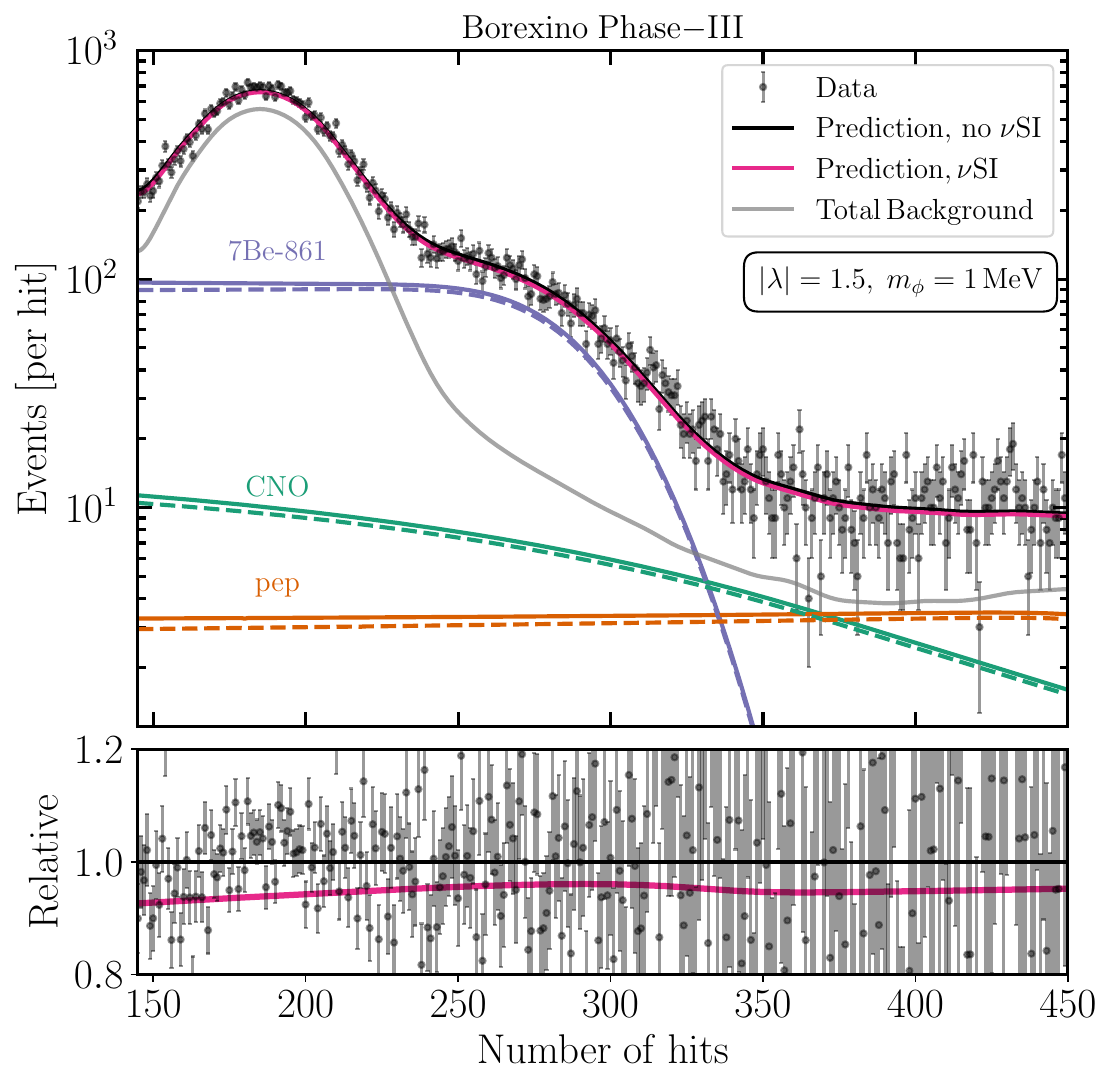}
\caption{Event spectrum of elastic electron–neutrino scattering as a function of the number of hits. Borexino Phase-III (subtracted) data are compared with the total predicted signal (black) and with the prediction including $\nu$SI (pink). Solid lines show the individual solar-neutrino components for the best-fit flux normalizations (Standard Model only), while dashed lines show a $\nu$SI scenario with $m_\phi = 10~\rm MeV$ and $\lambda = 1.5$. The total background is indicated by the gray line. In the lower panel, the relative deviation of the prediction including $\nu$SI from the total predicted signal is shown, with the total predicted signal normalized to 1.}
\label{fig:borexino_spectrum}
\end{figure}

The predicted recoil spectrum is obtained from
\begin{equation}
S^f_{\alpha i} = \mathcal{N}_e\mathcal{T}_{\rm live}\mathcal{\varepsilon}
\int_{N_h^{\,i}}^{N_h^{\,i+1}} \!\! \mathrm{~d}N_h
\int \mathrm{~d}T\, \frac{\mathrm{~d}\mathcal{R}}{\mathrm{~d}N_h}(T, N_h) 
\int_{E_\nu^{\min}(T)}^{E_\nu^{\max}} \! \mathrm{~d}E_\nu\,
\frac{\mathrm{~d}\phi_{e\alpha}^f}{\mathrm{~d}E_\nu}(E_\nu)
\frac{\mathrm{~d}\sigma_{\nu_\alpha e}}{\mathrm{~d}T}(E_\nu, T),
\end{equation}
where $E_\nu$ is the neutrino energy, $\mathrm{~d}\phi_{e\alpha}^f/\mathrm{~d}E_\nu$ denotes the differential neutrino flux of flavor $\alpha$ arriving at Earth after oscillations, originating from the solar source $f$, and we pick $f\in\{^7\mathrm{Be}, \mathrm{pep}, \mathrm{CNO}\}$. 
% \alpha$ the flavor arriving at Earth after oscillations
The detector response function \( \mathrm{~d}\mathcal{R}(T,N_h)/\mathrm{~d}N_h\) maps true recoil energy \(T\) to number of photomultiplier hits \(N_h\), modeled as Gaussian~\cite{Coloma:2022umy}.
For Phase-III, the normalization accounts for the number of electron targets $\mathcal{N}_e$ inside the 71.3 ton scintillator volume, the live time is $\mathcal{T}_{\rm live}=1431.6$ days, and the overall efficiency is $\mathcal{\varepsilon}=98.5\%$. 

Electron neutrinos produced in the solar core undergo flavor conversion via matter-induced effects, and later through vacuum oscillations in their journey to the Earth~\cite{Mikheyev:1985zog}.
We fix oscillation parameters to the NuFit best-fit values~\cite{Esteban:2020cvm}, assuming three-flavor evolution through the Sun, and adopt survival and appearance probabilities from Ref.~\cite{Mishra:2023jlq}.
\cref{fig:borexino_spectrum} shows the predicted spectra with and without $\nu$SI for $m_\phi=1~ \rm{MeV}$ and $|\lambda|=1.5$.
The effect of $\nu$SI is a suppression of the signal, with mild recoil-energy dependence.

\begin{figure}[t]
\centering
\includegraphics[width=0.8\textwidth]{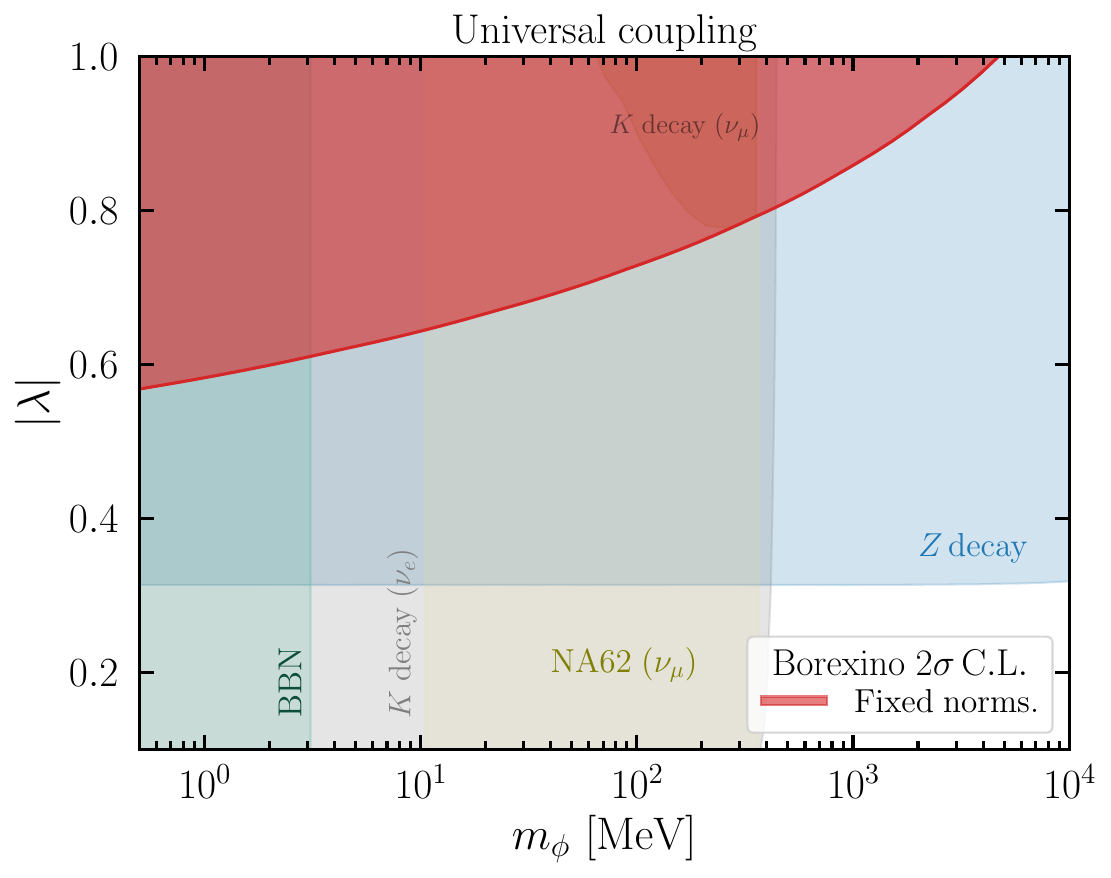}
\caption{Borexino bound on the parameter space of the neutrino self-interaction model. The red shaded region shows the $2\sigma$ confidence level obtained from an analysis of Borexino Phase III spectral data, with flux normalizations fixed to the Standard Solar Model predictions. For comparison, we also display constraints from the $Z$ invisible decay width discussed in Sec.~\ref{sec:Z}, as well as laboratory bounds from Kaon decay width and the NA62 shape analysis of kaon decays~\cite{Dev:2024ygx}, together with cosmological limits.}
\label{fig:borexino_2sigCL}
\end{figure}

To compute the allowed regions for $\nu$SI with Borexino, we follow an analysis similar to Ref.~\cite{Kelly:2024tvh}, but with the difference that, in addition to the $^7\rm Be$ flux, we also allow the $\rm pep$ and $\rm CNO$ flux normalizations to vary. The $\chi^2$ function used for the statistical analysis is given by
\begin{equation}
    \chi^2=\sum_i \frac{\left(N_i^{\text {obs}}-(B_i+S_i)\right)^2}{\sigma_i^2}+\chi_{\text {FN}}^2,
\end{equation}
where $N_i^{\text{obs}}$ is the observed spectrum, $\sigma_i$ are Borexino bin-by-bin uncertainties, and $\chi_{\text{FN}}^2$ enforces flux normalization priors. To ensure Gaussian statistics, every five bins are combined into one. We restrict the analysis to the photomultiplier hit range \(N_h \leq 450\), where \(N_h\) is an estimator of the recoil electron energy.
Above this region, the dominant background arises from ${}^{11}\mathrm{C}$~\cite{Borexino:2017rsf}.
All background rates $B_i$ are kept fixed throughout our analysis.
Our analysis uses the subtracted dataset, corresponding to \(64.28\%\) of the solar neutrino events.

The flux normalizations $\{n_{^7\mathrm{Be}},n_{\mathrm{pep}},n_{\mathrm{CNO}}\}$ here are defined relative to GS98~\cite{Grevesse:1998bj,Acharya:2024lke} SSM predictions; for CNO, all three components are rescaled simultaneously with fixed ratios.
We allow the flux normalizations to vary, subject to Gaussian priors implementing the solar luminosity constraint~\cite{Bergstrom:2016cbh}, as well as the $\rm pep$ to $\rm pp$ neutrino flux ratio, fixed by solar physics with a 2\% theoretical uncertainty.
We relegate the details of the analysis to~\cref{appB} -- therein (\cref{fig:borexino_norms}) we show the results of our Borexino analysis with respect to all free parameters (normalizations as well as new-physics parameters).

Here, we consider a relatively aggressive case, in which we fix the predicted flux normalizations to their SSM expectations. We then derive a constraint at $2\sigma$~CL as a function of $m_\phi$ and $\lambda$ as shown by the red region in~\cref{fig:borexino_2sigCL}. 
Even under this aggressive assumption, we see that the constraint from Borexino is weaker than other existing constraints, for instance that from the invisible decay width of the $Z$ boson. 
We explore how this constraint is further weakened if we allow the flux normalizations to vary in~\cref{fig:borexino_deltachi2}.
It is possible that with more precise solar-neutrino/electron scattering data, one can constrain self-interactions more precisely than other searches; however, such a search will be limited by systematic uncertainties associated with the flux normalizations.

For comparison, in~\cref{fig:borexino_2sigCL} we also show laboratory constraints from the kaon decay width and the NA62 shape analysis of kaon decays~\cite{Dev:2024ygx}, together with cosmological bound from the measurement of the effective number of neutrino species during the Big Bang Nucleosynthesis (BBN) era~\cite{Grohs:2020xxd}. Assuming $\Delta N_{\rm eff} \lesssim 0.3$~\cite{Escudero:2024uea}, the later translates into a lower bound of $m_\phi \gtrsim 3~\mathrm{MeV}$ for a complex scalar mediator that equilibrates with neutrinos before weak interaction decoupling.
%The results of the analysis of varying flux normalizations are shown in Fig.~\ref{fig:borexino_deltachi2}, presenting one- and two-dimensional $\Delta \chi^2$ projections. After marginalizing over flux normalization parameters, we derive the corresponding $1\sigma$ exclusion regions for $\nu$SI parameters as displayed in Fig.~\ref{fig:borexino_norms}. We also repeat the analysis with flux normalizations fixed to their SSM values. This yields stronger bounds, with the $1\sigma$ and $2\sigma$ confidence region shown in Fig.~\ref{fig:borexino_norms}.

\section{Discussion}\label{sec:discussion}

In this work, we present detailed calculations of radiative correction to neutrino electroweak couplings at one-loop level from novel neutrino self-interactions mediated by a light scalar particle $\phi$.
Our results, obtained by UV-completing the $\nu\nu\phi$ simplified operator in an extension of the Standard Model with an $SU(2)_L$ scalar triplet, are essential for making precision predictions for various neutrino-matter scattering processes. 
\begin{itemize}
\item We quantify their impact on the weak mixing angle measurements using $\nu$DIS and CE$\nu$NS processes. 
\item We refine the calculation of the $Z$-boson invisible decay width. 
\item For the case of $\nu_\tau$-philic neutrino self-interactions, we explore the effect on NSI between neutrino and matter and derive a leading constraint on the model parameter space for $m_\phi$ between 20\,MeV and GeV scales.
\item We perform a detailed study of solar-neutrino/electron scattering in Borexino. While the Borexino limit is still weaker than other existing constraints, our analysis paves the way for future improvement with larger-exposure and lower-noise detectors.
\end{itemize}
The breadth of these different processes demonstrates the wide-ranging consequences of invoking neutrino self-interactions beyond the SM.
Only by considering a global perspective can we fully test whether such self-interactions exist.
Our work points towards a number of future directions for further exploration, and potentially, discovery.

\bigskip
\textbf{Acknowledgments --} We thank Ayres Freitas and Ryan Plestid for useful discussions. SF and YZ are supported by a Subatomic Physics Discovery Grant (individual) from the Natural Sciences and Engineering Research Council of Canada. KJK is supported in part by US DOE Award \#DE-SC0010813.
This research was enabled in part by support provided by the Digital Research Alliance of Canada (\url{https://alliancecan.ca}).

\newpage
\appendix
\section{Feynman rules in the UV complete model}\label{appA}

In this appendix, we present the Feynman rules in the singlet ($\phi$) and triplet ($T$) extension of the Standard Model relevant for our calculations.
Following the main text, we define $g_Z^0 = g/(2\cos\theta_W)$, $g_W^0=g/\sqrt2$. 
The $Z$ boson is defined as $Z_\mu = \cos\theta_W W^3_\mu - \sin\theta_W B_\mu$.
The Standard Model gauge Feynman rules for neutrinos and charged leptons are
\begin{align}
&\begin{tikzpicture}
\begin{feynman}
\vertex [label=left:\(Z\)] (a) at (0,0);
\vertex (b) at (0.7,0);
\vertex [label=right:\(\nu_\alpha\)] (c) at (1.57,0.5);
\vertex [label=right:\(\nu_\beta\)] (d) at (1.57,-0.5);
\feynmandiagram [inline=(a.base)] {
(a)  -- [photon] (b),
(b) -- [fermion] (c),
(d) -- [fermion] (b),
};
\end{feynman}
\end{tikzpicture} \hspace{2cm} \raisebox{0.4cm}{$-i g_Z^0 \gamma^\mu \mathbb{P}_L \delta_{\alpha\beta}$} \nonumber \\[1ex]
&\begin{tikzpicture}
\begin{feynman}
\vertex [label=left:\(Z\)] (a) at (0,0);
\vertex (b) at (0.7,0);
\vertex [label=right:\(\nu_\alpha^c\)] (c) at (1.57,0.5);
\vertex [label=right:\(\nu_\beta^c\)] (d) at (1.57,-0.5);
\feynmandiagram [inline=(a.base)] {
(a)  -- [photon] (b),
(b) -- [fermion] (c),
(d) -- [fermion] (b),
};
\end{feynman}
\end{tikzpicture} \hspace{2cm} \raisebox{0.4cm}{$i g_Z^0 \gamma^\mu \mathbb{P}_R \delta_{\alpha\beta}$} \nonumber \\[1ex]
&\begin{tikzpicture}
\begin{feynman}
\vertex [label=left:\(W^-\)] (a) at (0,0);
\vertex (b) at (0.7,0);
\vertex [label=right:\(\ell^-_\alpha\)] (c) at (1.57,0.5);
\vertex [label=right:\(\nu_\beta\)] (d) at (1.57,-0.5);
\feynmandiagram [inline=(a.base)] {
(a)  -- [charged boson] (b),
(b) -- [fermion] (c),
(d) -- [fermion] (b),
};
\end{feynman}
\end{tikzpicture} \hspace{2cm} \raisebox{0.4cm}{$-i g_W^0 \gamma^\mu \mathbb{P}_L \delta_{\alpha\beta}$} \nonumber
\end{align} 
where $\alpha, \beta$ are used to denote lepton flavors.

Adding the $SU(2)_L$ triplet $T$ and turning on the $\phi$-$T^0$ mixing $\theta$, the corresponding gauge interactions are
\begin{align}
&\begin{tikzpicture}
\begin{feynman}
\vertex [label=left:\(Z\)] (a) at (0,0);
\vertex (b) at (0.7,0);
\vertex [label=right:\(T^0\)] (c) at (1.57,0.5);
\vertex [label=right:\(T^0\)] (d) at (1.57,-0.5);
\feynmandiagram [inline=(a.base)] {
(a)  -- [photon] (b),
(b) -- [charged scalar, momentum=\(p_1\)] (c),
(d) -- [charged scalar, momentum=\(p_2\)] (b),
};
\end{feynman}
\end{tikzpicture} \hspace{2cm} \raisebox{0.9cm}{$-2 i g_Z^0 (p_1+p_2)^\mu \cos^2\theta$} \nonumber\\
&\begin{tikzpicture}
\begin{feynman}
\vertex [label=left:\(Z\)] (a) at (0,0);
\vertex (b) at (0.7,0);
\vertex [label=right:\(T^0\)] (c) at (1.57,0.5);
\vertex [label=right:\(\phi\)] (d) at (1.57,-0.5);
\feynmandiagram [inline=(a.base)] {
(a)  -- [photon] (b),
(b) -- [charged scalar, momentum=\(p_1\)] (c),
(d) -- [charged scalar, momentum=\(p_2\)] (b),
};
\end{feynman}
\end{tikzpicture} \hspace{2cm} \raisebox{0.9cm}{$-2 i g_Z^0 (p_1+p_2)^\mu \cos\theta \sin\theta$} \nonumber\\
&\begin{tikzpicture}
\begin{feynman}
\vertex [label=left:\(Z\)] (a) at (0,0);
\vertex (b) at (0.7,0);
\vertex [label=right:\(\phi\)] (c) at (1.57,0.5);
\vertex [label=right:\(T^0\)] (d) at (1.57,-0.5);
\feynmandiagram [inline=(a.base)] {
(a)  -- [photon] (b),
(b) -- [charged scalar, momentum=\(p_1\)] (c),
(d) -- [charged scalar, momentum=\(p_2\)] (b),
};
\end{feynman}
\end{tikzpicture} \hspace{2cm} \raisebox{0.9cm}{$-2 i g_Z^0 (p_1+p_2)^\mu \cos\theta \sin\theta$} \nonumber\\
&\begin{tikzpicture}
\begin{feynman}
\vertex [label=left:\(W^-\)] (a) at (0,0);
\vertex (b) at (0.7,0);
\vertex [label=right:\(T^-\)] (c) at (1.57,0.5);
\vertex [label=right:\(T^0\)] (d) at (1.57,-0.5);
\feynmandiagram [inline=(a.base)] {
(a)  -- [charged boson] (b),
(b) -- [charged scalar, momentum=\(p_1\)] (c),
(d) -- [charged scalar, momentum=\(p_2\)] (b),
};
\end{feynman}
\end{tikzpicture} \hspace{2cm} \raisebox{0.9cm}{$ \sqrt{2} i g_W^0 (p_1+p_2)^\mu \cos\theta$} \nonumber \\
&\begin{tikzpicture}
\begin{feynman}
\vertex [label=left:\(W^-\)] (a) at (0,0);
\vertex (b) at (0.7,0);
\vertex [label=right:\(T^-\)] (c) at (1.57,0.5);
\vertex [label=right:\(\phi\)] (d) at (1.57,-0.5);
\feynmandiagram [inline=(a.base)] {
(a)  -- [charged boson] (b),
(b) -- [charged scalar, momentum=\(p_1\)] (c),
(d) -- [charged scalar, momentum=\(p_2\)] (b),
};
\end{feynman}
\end{tikzpicture} \hspace{2cm} \raisebox{0.9cm}{$ \sqrt{2} i g_W^0 (p_1+p_2)^\mu \sin\theta$} \nonumber 
\end{align} 

The Yukawa interactions of the scalars with neutrino and charged leptons are
\begin{align}
&\begin{tikzpicture}
\begin{feynman}
\vertex [label=left:\(T^0\)] (a) at (0,0);
\vertex (b) at (0.7,0);
\vertex [label=right:\(\nu_\alpha\)] (c) at (1.57,0.5);
\vertex [label=right:\(\nu_\beta^c\)] (d) at (1.57,-0.5);
\feynmandiagram [inline=(a.base)] {
(a)  -- [charged scalar] (b),
(b) -- [fermion] (c),
(d) -- [fermion] (b),
};
\end{feynman}
\end{tikzpicture} \hspace{2cm} \raisebox{0.4cm}{$i y_{\alpha\beta} \cos\theta$} \nonumber \\[1ex]
&\begin{tikzpicture}
\begin{feynman}
\vertex [label=left:\(T^0\)] (a) at (0,0);
\vertex (b) at (0.7,0);
\vertex [label=right:\(\nu_\alpha^c\)] (c) at (1.57,0.5);
\vertex [label=right:\(\nu_\beta\)] (d) at (1.57,-0.5);
\feynmandiagram [inline=(a.base)] {
(b)  -- [charged scalar] (a),
(b) -- [fermion] (c),
(d) -- [fermion] (b),
};
\end{feynman}
\end{tikzpicture} \hspace{2cm} \raisebox{0.4cm}{$i y_{\alpha\beta}^* \cos\theta$} \nonumber \\[1ex]
&\begin{tikzpicture}
\begin{feynman}
\vertex [label=left:\(\phi\)] (a) at (0,0);
\vertex (b) at (0.7,0);
\vertex [label=right:\(\nu_\alpha\)] (c) at (1.57,0.5);
\vertex [label=right:\(\nu_\beta^c\)] (d) at (1.57,-0.5);
\feynmandiagram [inline=(a.base)] {
(a)  -- [charged scalar] (b),
(b) -- [fermion] (c),
(d) -- [fermion] (b),
};
\end{feynman}
\end{tikzpicture} \hspace{2cm} \raisebox{0.4cm}{$i y_{\alpha\beta} \sin\theta = i \lambda_{\alpha\beta}$} \nonumber \\[1ex]
&\begin{tikzpicture}
\begin{feynman}
\vertex [label=left:\(\phi\)] (a) at (0,0);
\vertex (b) at (0.7,0);
\vertex [label=right:\(\nu_\alpha^c\)] (c) at (1.57,0.5);
\vertex [label=right:\(\nu_\beta\)] (d) at (1.57,-0.5);
\feynmandiagram [inline=(a.base)] {
(b)  -- [charged scalar] (a),
(b) -- [fermion] (c),
(d) -- [fermion] (b),
};
\end{feynman}
\end{tikzpicture} \hspace{2cm} \raisebox{0.4cm}{$i y_{\alpha\beta}^* \sin\theta = i \lambda^*_{\alpha\beta}$} \nonumber \\[1ex]
&\begin{tikzpicture}
\begin{feynman}
\vertex [label=left:\(T^-\)] (a) at (0,0);
\vertex (b) at (0.7,0);
\vertex [label=right:\(\nu_\alpha\)] (c) at (1.57,0.5);
\vertex [label=right:\(\ell^+_\beta\)] (d) at (1.57,-0.5);
\feynmandiagram [inline=(a.base)] {
(a)  -- [charged scalar] (b),
(b) -- [fermion] (c),
(d) -- [fermion] (b),
};
\end{feynman}
\end{tikzpicture} \hspace{2cm} \raisebox{0.4cm}{$- \frac{i}{\sqrt2} y_{\alpha\beta}$} \nonumber \\[1ex]
&\begin{tikzpicture}
\begin{feynman}
\vertex [label=left:\(T^-\)] (a) at (0,0);
\vertex (b) at (0.7,0);
\vertex [label=right:\(\ell^+_\alpha\)] (c) at (1.57,0.5);
\vertex [label=right:\(\nu_\beta\)] (d) at (1.57,-0.5);
\feynmandiagram [inline=(a.base)] {
(b)  -- [charged scalar] (a),
(b) -- [fermion] (c),
(d) -- [fermion] (b),
};
\end{feynman}
\end{tikzpicture} \hspace{2cm} \raisebox{0.4cm}{$- \frac{i}{\sqrt2} y^*_{\alpha\beta}$} \nonumber \\[1ex]
&\begin{tikzpicture}
\begin{feynman}
\vertex [label=left:\(T^{--}\)] (a) at (0,0);
\vertex (b) at (0.7,0);
\vertex [label=right:\(\ell^-_\alpha\)] (c) at (1.57,0.5);
\vertex [label=right:\(\ell^+_\beta\)] (d) at (1.57,-0.5);
\feynmandiagram [inline=(a.base)] {
(a)  -- [charged scalar] (b),
(b) -- [fermion] (c),
(d) -- [fermion] (b),
};
\end{feynman}
\end{tikzpicture} \hspace{2cm} \raisebox{0.4cm}{$-i y_{\alpha\beta}$} \nonumber \\[1ex]
&\begin{tikzpicture}
\begin{feynman}
\vertex [label=left:\(T^{--}\)] (a) at (0,0);
\vertex (b) at (0.7,0);
\vertex [label=right:\(\ell^+_\alpha\)] (c) at (1.57,0.5);
\vertex [label=right:\(\ell^-_\beta\)] (d) at (1.57,-0.5);
\feynmandiagram [inline=(a.base)] {
(b)  -- [charged scalar] (a),
(b) -- [fermion] (c),
(d) -- [fermion] (b),
};
\end{feynman}
\end{tikzpicture} \hspace{2cm} \raisebox{0.4cm}{$-i y^*_{\alpha\beta}$} \nonumber
\end{align}

\section{Borexino analysis}\label{appB}
In our analysis, we define the reduced solar flux normalization parameters as
\begin{equation}
n_i = \frac{\Phi_i}{\Phi_i^{\rm SSM}}, \quad i = {}^7\mathrm{Be}, \ \mathrm{pep}, \ \mathrm{CNO},
\end{equation}
where $\Phi_i^{\rm SSM}$ are the reference neutrino fluxes from the latest GS98 high metallicity Standard Solar Model predictions~\cite{Vinyoles:2016djt,Acharya:2024lke}.

\begin{figure}[t]
\centering
\includegraphics[width=0.8\textwidth]{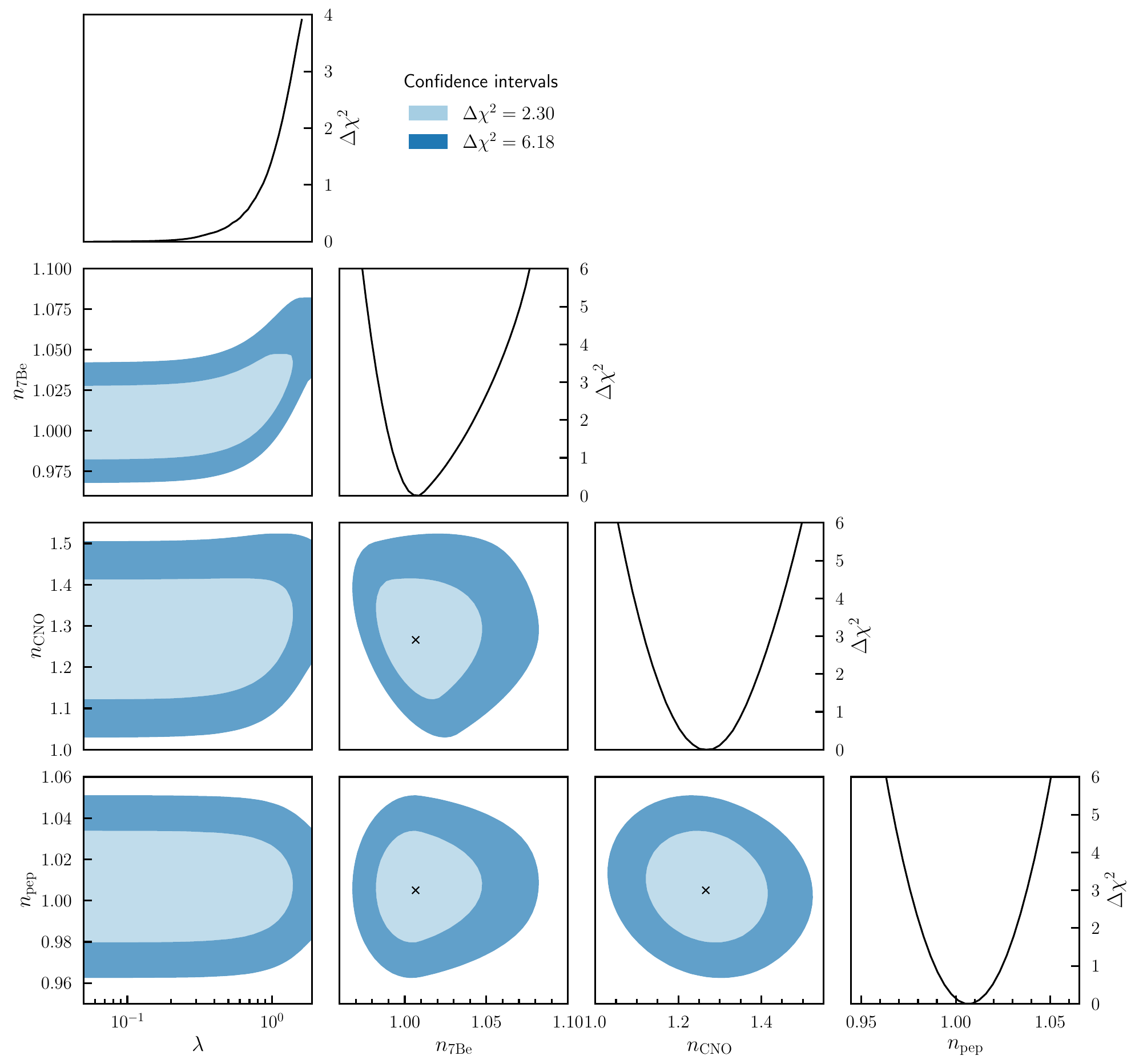}
\caption{Borexino Phase III analysis of the neutrino self-interaction universal coupling for a fixed mediator mass ($m_{\phi}=10~\rm MeV$), allowing the $^7\rm Be$, $\rm pep$, and $\rm CNO$ solar neutrino flux normalizations to vary with Gaussian priors around their Standard Solar Model (SSM) predictions. The off-diagonal panels show the two-dimensional allowed regions at 1$\sigma$ and 2$\sigma$ confidence levels (for 2 d.o.f., with corresponding intervals $\Delta\chi^2 = 2.30, 6.18$), while the diagonal panels display the one-dimensional $\Delta\chi^2$ profiles for each parameter. In all cases, results are obtained after marginalizing over the parameters not shown. Best-fit points for the fluxes are marked with an “x,” and $n_i=1$ corresponds to the SSM values.}
\label{fig:borexino_deltachi2}
\end{figure}

When the solar flux normalizations are allowed to vary freely, following Ref.~\cite{Gonzalez-Garcia:2023kva}, we impose priors from solar luminosity constraints~\cite{Bahcall:2001pf}. These constraints require that the total thermal energy generated, together with each solar neutrino flux, matches the observed solar luminosity. The corresponding $\chi^2$ contribution is
\begin{equation}
\chi_{\rm LC}^2 = \frac{\left(\sum_i \beta_i (1 - n_i)\right)^2}{(0.0034)^2},
\end{equation}
where $\beta_i$ denotes the fractional contribution of the $i$th nuclear reaction to the total solar luminosity, with numerical values taken from~\cite{Gonzalez-Garcia:2023kva}. The pep to pp neutrino flux ratio is predicted with very small theoretical uncertainty, and we implement this constraint by imposing a 2$\%$ Gaussian prior on $f_{\text{pep}}$, reflecting the spread among different SSM predictions. For the CNO fluxes, the three components are scaled simultaneously by a single normalization parameter while keeping their relative ratios fixed as predicted by the SSM.

\begin{figure}[t]
\centering
\includegraphics[width=0.8\textwidth]{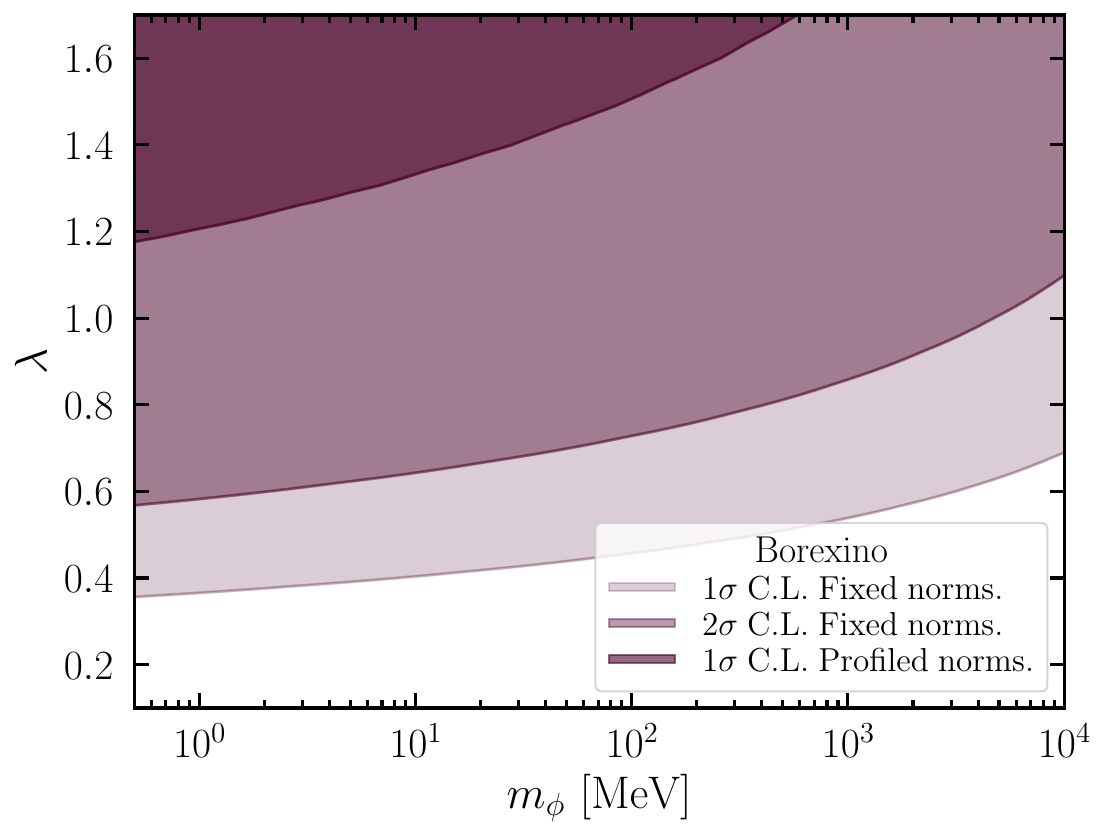}
\caption{Constraints on neutrino self-interaction parameters derived from Borexino measurements of the solar neutrino fluxes. Shaded regions show two scenarios: (i) flux normalizations fixed to the Standard Solar Model (GS98) predictions, yielding the $1\sigma$ and $2\sigma$ confidence regions, and (ii) flux normalizations varied subject to priors, which leads to a weaker bound, with the marginalized $1\sigma$ exclusion region indicated in dark purple.}
\label{fig:borexino_norms}
\end{figure}

In the analysis allowing the solar flux normalizations to vary, we scan over all flux parameters together with the $\nu$SI parameters $(|\lambda|, m_\phi)$. The resulting one- and two-dimensional $\Delta \chi^2$ projections are presented in Fig.~\ref{fig:borexino_deltachi2} for a fixed mediator mass of $m_{\phi}=10~\rm MeV$. In the presence of neutrino self-interactions, the 1$\sigma$ solar flux normalizations, after profiling over all other parameters including $|\lambda|$ and $m_\phi$, are determined to be
\begin{equation}
n_{{}^7\rm Be} = 1.008_{-0.024}^{+0.041}, \quad n_{\rm pep} = 1.005_{-0.025}^{+0.029}, \quad n_{\rm CNO} = 1.266 _{-0.14}^{+0.15}.
\end{equation}
We show the corresponding $1\sigma$ confidence region for the $\nu$SI parameters in Fig.~\ref{fig:borexino_norms}, obtained by marginalizing over the solar flux normalizations.

A more rigorous limit can be achieved by repeating the analysis with the flux normalizations fixed to their SSM values, and the corresponding $1\sigma$ and $2\sigma$ confidence regions for this fixed-normalization analysis are shown in Fig.~\ref{fig:borexino_norms}.

\bibliographystyle{utphys}
\bibliography{Refrences}

\end{document}